\documentclass[twocolumn]{aastex631}

\defcitealias{MayKomacek2021}{May \& Komacek et al.}

\received{October 29, 2021}
\revised{March 18, 2022}
\accepted{-}

\submitjournal{ApJ}

\shorttitle{gCMCRT}
\shortauthors{Lee et al.}
\graphicspath{{./}{plots/}}

\begin{document}

\title{3D radiative-transfer for exoplanet atmospheres. gCMCRT: a GPU accelerated MCRT code.}

\correspondingauthor{Elspeth K.H. Lee}
\email{elspeth.lee@csh.unibe.ch}

\author[0000-0002-3052-7116]{Elspeth K.H. Lee}
\affiliation{Center for Space and Habitability, University of Bern, Gesellschaftsstrasse 6, CH-3012 Bern, Switzerland}

\author{Joost P. Wardenier}
\affiliation{Atmospheric, Oceanic \& Planetary Physics, Department of Physics, University of Oxford, Oxford OX1 3PU, UK}

\author{Bibiana Prinoth}
\affiliation{Lund Observatory, Department of Astronomy and Theoretical Physics, Lunds Universitet, Solvegatan 9, 222 24 Lund, Sweden}

\author{Vivien Parmentier}
\affiliation{Atmospheric, Oceanic \& Planetary Physics, Department of Physics, University of Oxford, Oxford OX1 3PU, UK}

\author{Simon L. Grimm}
\affiliation{Center for Space and Habitability, University of Bern, Gesellschaftsstrasse 6, CH-3012 Bern, Switzerland}

\author{Robin Baeyens}
\affiliation{Institute of Astronomy, KU Leuven, Celestijnenlaan 200D, 3001, Leuven, Belgium}

\author{Ludmila Carone}
\affiliation{Max-Planck-Institut f{\"u}r Astronomie, K{\"o}nigstuhl 17, D-69117 Heidelberg, Germany}

\author{Duncan Christie}
\affiliation{Astrophysics Group, University of Exeter, Stocker Road, Exeter, EX4 4QL, UK}

\author{Russell Deitrick}
\affiliation{Center for Space and Habitability, University of Bern, Gesellschaftsstrasse 6, CH-3012 Bern, Switzerland}

\author{Daniel Kitzmann}
\affiliation{Center for Space and Habitability, University of Bern, Gesellschaftsstrasse 6, CH-3012 Bern, Switzerland}

\author{Nathan Mayne}
\affiliation{Astrophysics Group, University of Exeter, Stocker Road, Exeter, EX4 4QL, UK}

\author[0000-0002-0786-7307]{Michael Roman}
\affiliation{School of Physics and Astronomy, University of Leicester, University Road, Leicester LE1 7RH, UK}

\author{Brian Thorsbro}
\affiliation{Lund Observatory, Department of Astronomy and Theoretical Physics, Lunds Universitet, Solvegatan 9, 222 24 Lund, Sweden}



\begin{abstract}
Radiative-transfer (RT) is a key component for investigating atmospheres of planetary bodies.
With the 3D nature of exoplanet atmospheres being important in giving rise to their observable properties, accurate and fast 3D methods are required to be developed to meet future multi-dimensional and temporal data sets.
We develop an open source GPU RT code, gCMCRT, a Monte Carlo RT forward model for general use in planetary atmosphere RT problems.
We aim to automate the post-processing pipeline, starting from direct global circulation model (GCM) output to synthetic spectra.
We develop albedo, emission and transmission spectra modes for 3D and 1D input structures.
We include capability to use correlated-k and high-resolution opacity tables, the latter of which can be Doppler shifted inside the model.
We post-process results from several GCM groups including ExoRad, SPARC/MITgcm THOR, UK Met Office UM, Exo-FMS and the Rauscher model.
Users can therefore take advantage of desktop and HPC GPU computing solutions.
gCMCRT is well suited for post-processing large GCM model grids produced by members of the community and for high-resolution 3D investigations.
\end{abstract}



\section{Introduction}
\label{sec:intro}

The volume of observational data of exoplanet atmospheres from telescopes continues to grow exponentially, uncovering a wide variety of atmospheric properties across the exoplanet population.
With the upcoming launch of JWST \citep{Bean2018} and the Ariel mission \citep{Tinetti2016}, precise spectral data across the infrared wavelength regime will become available for the community to analyse.
Kepler \citep{Howell2014}, TESS \citep{Ricker2014} and CHEOPS \citep{Broeg2013} have provided optical band photometric data of exoplanet phase curves, allowing examination of their albedo and thermal properties \citep[e.g.][]{Demory2013, Wong2020, Morris2021}.
The unique 3D nature of each exoplanet atmosphere manifests itself in the observational properties of that specific object.
Therefore, a 3D understanding of their atmospheres is key to a holistic investigation of their atmospheric structure and composition.
For phase curve observations, different hemispheres of the planet come in and out of view of the observational direction, leading to an intrinsically 3D mapping of the atmosphere.
For the dayside of atmospheres that are highly irradiated and puffed up, for example, ultra hot Jupiters, the radiation from the dayside can be blended with the cold nightside emission when both the dayside and nightside hemispheres are in view, for example when taking phase curve observations \citep[e.g.][]{Irwin2020,Taylor2021}.
These close-in planets are also expected to be tidally locked to their host star, showing a permanent dayside and nightside hemisphere.
In addition, as result of the 3D dynamical properties of these UHJ planets and their short dayside radiative timescales, their day-to-night energy transport can be weak \citep[e.g.][]{Parmentier2018}.
Accounting for all these effects accurately requires 3D RT modelling techniques, able to produce spectra taking into account the contribution from the entire 3D atmosphere to the end result.

In the past decade, the number and precision of high resolution observations has greatly increased, revealing the rich chemical composition of hot Jupiter atmospheres \citep[e.g.][]{Seidel2019, Ehrenreich2020, Merritt2020, Pino2020}.
Current high-resolution spectral instruments, for example, ESPRESSO \citep{Pepe2021}, CARMENES \citep{Quirrenbach2016}, GIANO-B \citep{Oliva2012} and CRIRES+ \citep{Follert2014} provide detailed mapping of the chemical structure and wind profiles in the atmosphere.
The fidelity and wavelength range of such high-resolution data will also increase as the next-generation of telescopes and instruments come online in the next decade such as HIRES on the EELT \citep{Marconi2016}.
The number of wavelength points for these high-resolution studies is large.
For example, there are approximately 170,000 wavelength points in the CRIRES+ wavelength range (0.95 $\mu$m $<$ $\lambda$ $<$ 5.2 $\mu$m) for a resolution of R = 100,000.
Performing RT for 3D models with this many wavelength points is a challenging prospect, not withstanding the fact that modelling at high resolution is typically performed at many times the resolution of the spectrograph.
Should phase information be required, this RT modelling needs to be repeated for each phase of interest.
For modellers, there is a need for fast, accurate and flexible 3D model post-processing capabilities that are able to meet the challenges of modelling planets at high-resolution.

As the number of characterised planets increases in both low and high-resolution, with 1000s of exoplanets scheduled to be characterised in the coming decades, interpreting physical trends from the observational data may become possible.
However, this requires large parameter grids of models to be run, which can be a time consuming and expensive undertaking.
Several modelling groups have performed parameter investigations using GCMs, from examining effects of equilibrium temperature, atmospheric drag and chemistry \citep{Komacek2017, Komacek2019,Baeyens2021} to grids of models used to examine cloud properties across the hot Jupiter equilibrium regime \citep{Parmentier2016, Parmentier2021, Roman2021}.
Post-processing these large grids is a significant project in itself, typically requiring processing simulations manually one-by-one.
Therefore, typically modellers use available data from the GCM model itself such as the temperature structures and outgoing longwave radiation data to make large scale diagnostic prescriptions.
For future large GCM grid projects, there is a need for a uniform and fast spectral 3D RT processing method, able to process the raw GCM output formats and perform the RT modelling in a simple and automated manner.

Recently, several studies have investigated biases that occur when the 3D structure of the atmosphere is not taken into account.
In emission, \citet{Feng2016} and \citet{Taylor2021} investigate the biases that can occur when retrieving emission spectra assuming a single T-p profile across a hemisphere that is inhomogeneous in temperature.
They find in some cases that assuming a single global T-p profile can produce spurious molecular detections from the blending of cooler and hotter profiles of the atmosphere.
\citet{Blecic2017} retrieved 1D T-p profiles of 3D GCM output from \citet{Dobbs-Dixon2013}.
They found that it was challenging to represent the 3D T-p structure dayside well using 1D retrieved T-p profiles.
\citet{Irwin2020} and \citet{Feng2020} present 2.5D retrieval methods that are able to take into account the variable T-p structures of the atmosphere, increasing accuracy when retrieving phase curve data.
\citet{Dobbs-Dixon2017} show how the wavelength dependent photosphere changes across the planet due to changes in the temperature and chemical composition, showing a complex contribution function structure that varies across the globe.
\citet{Fortney2019} investigate this photospheric radius effect, suggesting that not taking into account the 3D emitting area of the planet can lead to errors of 5\% or more in the flux.

In transmission, \citet{Caldas2019} and \citet{Pluriel2020} investigated biases that occur when not assuming 3D transmission geometry, showing significant differences between the full 3D model and a mean profile and that retrievals may produce biased results if the 3D chemical and temperature structure of the atmosphere is not taken into account.
\citet{Line2016b} show that non-uniform, patchy cloud structures between the east and west terminators can be degenerate with high molecular weight.
\citet{MacDonald2020} and \citet{Lacy2020} have also shown that chemical gradients in the 3D structure across both east and west limbs can also produce biased results when retrieved in 1D.

To date, modelling high-resolution spectra in 3D has mostly stemmed from the \citet{Kempton2012} model, with the exception of the \citet{Showman2013c} study.
\citet{Flowers2019} post-process 3D hot Jupiter GCMs to investigate the high-resolution transmission spectra and their cross-correlation properties.
\citet{Beltz2021} show that post-processing using the 3D structure of GCM models improves the detection significance of molecules at high resolution in emission.
Both of these studies used output from a \citet{Rauscher2012} GCM model.
\citet{Savel2021} use MITgcm models of WASP-76b from \citetalias{MayKomacek2021} (\citeyear{MayKomacek2021}) to produce high-resolution spectra, finding high-altitude cloud particles can reproduce well the phase variations observed by \citet{Ehrenreich2020}.
\citet{Wardenier2021} use SPARC/MITgcm \citep{Showman2009} GCM output of a WASP-76b model, finding Fe depletion at the limbs of the planet can also provide a reasonable explanation for the \citet{Ehrenreich2020} data.

From the above studies, it is highly warranted to develop the accurate and flexible 3D radiative-transfer techniques required to tackle the challenges associated with the upcoming high fidelity observational data on exoplanet atmospheres.
These requirements make the Monte Carlo Radiative Transfer (MCRT) method, a highly suitable methodology to meet these demands as it is able to directly take into account the 3D geometry of the atmosphere, include multiple scattering effects and other photon microphysics.
In transmission, there is a need for a model able to take into account both the chemical and cloud gradients in a 3D, inhomogeneous manner, weighting correctly the contribution of each transmission limb to the transmission spectra.
In emission, a model must now take into account the inhomogeneous nature of the 3D temperature structure and correctly weight each emitting area of the exoplanet.

MCRT is a standard 3D RT method in the astrophysical community \citep{Steinacker2013, Noebauer2019}, but has seen limited use in planetary contexts \citep[e.g.][]{Hood2008,deKok2012,Munoz2015,Robinson2017,Stolker2017}.
MCRT methods have also been used for simulations of RT through Earth clouds \citep[e.g.][]{Mayer2009}.
In astrophysics, MCRT models have been used to post-process hydrodynamical models to produce synthetic spectra and compare to observation data \citep[e.g.][]{Robitaille2011}.
Additionally, directly using MCRT as the RT solver inside the hydrodynamical model itself has been performed, for example in the protoplanetary disk modelling of \citet{Harries2019}.
3D GCM model output is typically transformed into latitude-longitude grids, making 3D spherical geometry a natural choice for the MCRT model.
Of primary interest is producing transmission, emission and albedo spectra, as well as phase curves.
Enabling the calculation of 3D contribution functions is also important to establish what sections of the planet are contributing to the observable features.

In this study, we develop gCMCRT, an open source GPU version of the \citet{Lee2017, Lee2019} CMCRT model\footnote{https://github.com/ELeeAstro} for general use in planetary atmosphere radiative-transfer.
CMCRT has been used to post-process GCM simulations from \citet{Lee2016}, \citet{Lee2020} and \citet{Lee2021}, showing its usefulness in 3D contexts.
In \citet{Wardenier2021}, a high wavelength resolution version was developed, including the Doppler shift of the local opacity from rotation and winds, to investigate the cross-correlation signal of Fe in transmission of WASP-76b \citep{Ehrenreich2020,Kesseli2021}, demonstrating the importance of considering the 3D structure of the atmosphere for high-resolution interpretation as well.
This used SPARC/MITgcm model output \citep{Showman2009} as input to CMCRT.

By applying GPU technology to CMCRT it is now easier and faster to perform 3D modelling without the need of high performance CPU environments.
Due to the high level of parallelism of MCRT methods, the GPU offers a speed up of 3-10x over the older CPU version.
This current study was performed solely on a desktop computer with a Nvidia RTX 3080 GPU card.

In this paper, we develop several standard modes for gCMCRT:
\begin{enumerate}
\item Albedo spectra
\item Transmission spectra
\item Emission spectra
\item Hi-res transmission spectra
\item Hi-res emission spectra
\end{enumerate}
The albedo and emission modes are able to model phase curves by performing the experiment across multiple viewing angles and at different planetary phases.
We outline the physics and mechanisms used in gCMCRT to calculate spectral properties of model atmospheres.
We post-process hot Jupiter GCMs using standard output from the \citet{Rauscher2010,Rauscher2012,Rauscher2013} model, SPARC/MITgcm \citep{Showman2009}, UK Met-Office UM \citep{Mayne2014}, ExoRad \citep{Carone2020}, THOR \citep{Mendonca2016, Deitrick2020}, and Exo-FMS \citep{Lee2021}.
Each model is used to showcase a mode of gCMCRT.

In Sect. \ref{sec:gCMCRT}, \ref{sec:ms} and \ref{sec:varred} we detail how the MCRT techniques and physics are performed in gCMCRT.
Sect. \ref{sec:obs} gives how observable quantities are derived from the gCMCRT output.
Sect. \ref{sec:confunc} shows how 3D contribution functions are calculated for the albedo and emission modes.
Sect. \ref{sec:opac} details the opacity sources that gCMCRT can use.
Sect. \ref{sec:CEab} details how chemical abundances are implemented in gCMCRT.
Sect. \ref{sec:albspec} shows the results of our 3D albedo spectra mode post-processing.
Sect. \ref{sec:emspec} shows the results of the emission spectra post-processing.
Sect. \ref{sec:transspec} shows the results of the transmission spectra post-processing.
Sect. \ref{sec:hiem} shows the results of the high-resolution emission spectra post-processing.
Sect. \ref{sec:hitrans} shows the results of the high-resolution transmission spectra post-processing.
Sect. \ref{sec:disc} and \ref{sec:con} present the discussion and conclusion of our results respectively.

\section{\lowercase{g}CMCRT}
\label{sec:gCMCRT}

gCMCRT is a Monte Carlo Radiative Transfer (MCRT) code that simulates the path of photon packets through a planetary atmosphere in three dimensions.
It is a hybrid MCRT and raytracing code, using the `peeloff' raytracing method (Sect. \ref{sec:peeloff}) to produce images and spectra of the simulated planet.
Both the basic MCRT and raytracing capabilities can be used separately or together, depending on the desired output.
Basic MCRT methods have been extensively detailed in many sources \citep[e.g.][]{Dupree2002, Whitney2011, Noebauer2019}.
gCMCRT primarily uses a spherical geometry grid, with packets moving and interacting inside the 3D computational domain.
As such it is sub-grid method, meaning that packets evolve through 3D space and the ray tracing is performed within the spherical geometry, not requiring special placement or selection of vertical layers.

gCMCRT is originally based on the model published by \citet{Hood2008}, which has its origins in MCRT codes used to investigate radiative-transfer and photo-ionisation for astrophysical applications \citep[e.g.][]{Wood2004}.
This model was expanded and specialised for exoplanetary atmosphere science in \citet{Lee2017} and benchmarked to contemporary 1D RT codes with correlated-k opacities in \citet{Lee2019}.
In this work, in addition to adapting to GPU architecture, we further expand the capabilities of the model, making it suitable for a wide variety of RT situations found in planetary science.

The user interacts with the  MCRT routines through a custom `main experiment' routine which the user can then use to call the core routines to perform a series of operations for the photon packets for their desired set-up.
We include standard set-ups for each mode to provide a baseline usability of the model.

MCRT is generally described as `embarrassingly parallel', with the program able to be relatively simply parallelised for multiple processors.
An advantage of utilising GPU cards is the large number (1000s) of cores on each GPU card compared to CPUs (10s).
This makes them ideal for simulating MCRT models where a large number of independent, simple calculations are required to be performed.
This allows the model to be run on desktop machines with GPU cards cheaply, whereas the CPU version required specialised HPC equipment to function efficiently.
GPU nodes have also become a popular addition to compute servers in recent years, providing magnitudes of computing scalability for models able to take advantage of the large number of cores.

gCMCRT is written in CUDA Fortran, chosen as it is generally a more familiar coding language to scientists in the astrophysics and (exo)planetary science fields, allowing the user a simpler designing of experiments to interface with the core routines.
CUDA Fortran retains the Fortran syntax and program layout with additional special syntax that enables communication with the GPU architecture.
This also allows future development of gCMCRT to take advantage of other well tested MCRT methods such as photo-ionisation \citep{Wood2013}.

We have written gCMCRT in a highly modular way, with the core routines modifying packet properties and performing the raytracing calculations.
Grid and opacity data are stored in global GPU modules, able to be accessed by routines that require them.
The host (CPU) side of the code reads and calculates the opacity structure of the atmosphere, which is then passed to the GPU memory.
Image data is also stored in global GPU memory, updated by each packet as it contributes to the image.
After the iteration is finished, the GPU data is given back to the host for output.

In testing, the main bottleneck was found to be the reading of 3D opacity data from the hard disk by the CPU.
We have optimised this section of code as best as possible, such as using asynchronous I/O and loading the maximum amount of data into memory as possible each read statement.

We have specifically avoided the use of additional packages inside the Fortran code (such as HDF5) to simplify the usability of the model.
Instead, output is produced in Fortran binary files, which can then be converted to more flexible storage formats later.

The main methods used in gCMCRT are already detailed in \citet{Lee2017} and \citet{Lee2019}.
However, in the following sections, we provide a brief review of the main processes and describe new capabilities included in gCMCRT.

\section{Multiple scattering}
\label{sec:ms}

gCMCRT includes the ability to treat multiple or single scattering using a variety of different phase functions.
Due to the flexibility of the method, mixing different scattering phase functions for varying scattering sources can also be performed.

\subsubsection{Isotropic scattering}
For isotropic scattering the phase function, $P(\cos\theta)$, is given by
\begin{equation}
P_{\rm Iso}(\cos\theta) = 1.
\end{equation}
Sampling a cosine angle for isotropic scattering is simply given by
\begin{equation}
\cos\theta =  2\zeta - 1,
\end{equation}
where $\zeta$ is a uniformly sampled random number between 0 and 1.
This equally samples a direction across 4$\pi$ steradians.

\subsubsection{Lambertian surface scattering}

For scattering of photon packets off a surface, a common and simple phase function to use is the Lambertian phase function
\begin{equation}
P_{\rm Lam}(\cos\theta) = \cos\theta,
\end{equation}
which equally samples a hemisphere of $\pi$ steradians.
Sampling a cosine angle for Lambertian scattering is given by
\begin{equation}
\cos\theta = \sqrt{\zeta}.
\end{equation}

\subsubsection{Rayleigh scattering}

For scattering off gas particles or small size parameter cloud particles, the Rayleigh scattering phase function is used, given by
\begin{equation}
P_{\rm Ray}(\cos\theta) = \frac{3}{4} (1 + \cos^{2}\theta),
\end{equation}
a cosine angle for unpolarised Rayleigh scattering is sampled through \citep[e.g.][]{Mayer2009, Frisvad2011}
\begin{equation}
\cos\theta = \sqrt[3]{-q + \sqrt{1 + q^{2}}}, \ q = 4\zeta - 1.
\end{equation}

\subsubsection{Aerosol scattering}

A key characteristic of planetary atmospheres is the presence of aerosol material, either photochemically produced hazes or cloud particles formed by condensation.
Accurate treatment of light scattering by aerosols is a vital part of RT and gCMCRT includes several options to treat aerosol scattering.

The Henyey-Greenstein (HG) phase function \citep{Henyey1941} has been extensively used as a one parameter fit to the Mie scattering phase function.
The HG function is given by
\begin{equation}
P_{\rm HG}(\cos\theta) = \frac{1-g^{2}}{(1+g^{2}-2g\cos\theta)^{3/2}},
\end{equation}
where $g$ = $\langle\cos\theta\rangle$, is the asymmetry factor defined as the mean cosine scattering angle.
Sampling a cosine angle from the HG function is given by \citep[e.g.][]{Sharma2015}
\begin{equation}
\label{eq:HG_samp}
\cos\theta = \frac{1}{2g}\left[1 + g^{2} - \left(\frac{1-g^{2}}{1 - g + 2g\zeta}\right)^{2}\right].
\end{equation}

One of the main limitations of the HG function is that as $g$ approaches zero, the phase function becomes isotropic rather than Rayleigh.
This generally occurs for IR wavelengths and for small particles at optical wavelengths.
Such particles make a significant proportion of the cloud size distributions in exosolar atmospheres \citep[e.g.][]{Powell2018}.
To address this, \citet{Draine2003} propose a hybrid HG and Rayleigh phase function with the form
\begin{equation}
\label{eq:Draine_P}
P_{\rm D03}(\cos\theta) = \frac{1-G^{2}}{1 + \alpha(1+2G^{2})/3}\frac{1+\alpha\cos^{2}\theta}{(1+G^{2}-2G\cos\theta)^{3/2}},
\end{equation}
where G($g$) is now a function of the asymmetry parameter $g$, assuming $\alpha$ is given \citep{Draine2003,Sharma2015}.
When $\alpha$ = 1, this distribution is equivalent to the the \citet{Cornette1992} function.
We note that in \citet{Draine2003} a better fit to the Mie scattering phase function was produced for interstellar dust when $\alpha$ $<$ 1, therefore we assume $\alpha$ = 0.5 as a default in CMCRT.
Sampling a cosine angle from Eq. \ref{eq:Draine_P} is performed following the Gibbs sampling approach found in \citet{Zhang2019}.

A variant of the HG function is the two term HG function (TTHG), a linear combination of two HG functions, given by
\begin{equation}
P_{\rm TTHG}(\cos\theta) = \alpha P_{\rm HG}(g_{1}) + (1-\alpha)P_{\rm HG}(g_{2}),
\end{equation}
with $\alpha$ representing the forwardly scattered fraction, $g_{1}$ $\geq$ 0 the forward scattered asymmetry parameter and $g_{2}$ $\leq$ 0 the backward scattered asymmetry parameter.
This attempts to better capture the backward scattering fraction produced by Mie calculations, especially important at optical wavelengths.
Sampling a cosine angle for the TTHG function is performed by sampling a random number.
Should $\zeta$ $<$ $\alpha$ the new direction is sampled using Eq. \ref{eq:HG_samp} with $g$ = $g_{1}$, otherwise Eq. \ref{eq:HG_samp} is sampled with $g$ = $g_{2}$ \citep{Pfeiffer2008}.

For the TTHG function, there is no known general solution to find the parameters $\alpha$, $g_{1}$ and $g_{2}$ from the asymmetry factor $g$ directly, though fitting to individual aerosol materials using higher moments of the TTHG function has been performed \citep[e.g.][]{Kattawar1975}.
As the default in CMCRT, we use the simple relationship used in \citet{Cahoy2010, Batalha2019}, which has been found useful for reproducing Jupiter reflectance spectra, these are given as: $g_{1}$ = $g$, $g_{2}$ = -$g$/2 and $\alpha$ = 1 - $g_{2}^{2}$.

An interesting consequence of the MCRT method is that the weighted asymmetry factor for the mixture of gas and cloud particles does not need to be calculated.
Instead, the probability of a scattering event off a gas particle, $P_{\rm gas}$, is sampled directly as
\begin{equation}
P_{\rm gas} = \frac{\kappa_{\rm Ray}}{\kappa_{\rm Ray} + \kappa_{\rm sca, cld}},
\end{equation}
where $\kappa_{\rm Ray}$ [cm$^{2}$ g$^{-1}$] is the Rayleigh scattering opacity and $\kappa_{\rm sca, cld}$ [cm$^{2}$ g$^{-1}$] is the scattering opacity of the cloud.
A random number can then be drawn every scattering event to determine the appropriate scattering phase function to sample.

\section{Variance reduction methods}
\label{sec:varred}

The act of including variance reduction schemes, also known as `biasing', attempts to address both the signal properties and variance issues present in a default MCRT scheme.
For 3D applications, some form of variance reduction is usually required beyond simple MCRT set-ups and is a key to retaining high accuracy and reducing noise in the model.
Below we briefly outline the variance reduction techniques currently implemented as options in gCMCRT.

\subsection{Continuous absorption}

In the default, non-biased MCRT methodology, the weight of the packet (representing the fraction of energy carried by the packet) is usually kept constant as it travels through the simulation until it interacts with the medium in some way.
In the continuous absorption method however, the weight of the packet degrades smoothly as it travels through the simulation, given by \citep[e.g.][]{Noebauer2019}
\begin{equation}
W_{\rm ph, new} = W_{\rm ph} e^{-\tau_{\rm abs}},
\end{equation}
where $W_{\rm ph, new}$ is the new weight of the packet, $W_{\rm ph}$ the original weight and $\tau_{\rm abs}$ the optical depth of the absorption opacity.
This smooths the weight reduction of the packets as they travel through the simulation and is mostly useful for when the atmosphere is highly absorbing.

\subsection{Forced interaction}

We include three forced interaction mechanisms, forced scattering \citep{Cashwell1959, Steinacker2013}, where the packet is forced to scatter within the bounds of the simulation, path length stretching \citep{Baes2016} and hybrid forced scattering with path length stretching \citep{Baes2016}.
These methods require an extra raytracing step to calculate the optical depth to the end of the grid in the direction of the packet path.
The packet weight is also increased or reduced given by the appropriate weighing function for each method.
By altering the calling structure of the experiment, each forced interaction method can also be called only once at the start of the packet integration, known as `forced first scattering'.

\subsection{Emission biasing}

In a non-biased MCRT scheme, the number of packets simulated in emission for each cell, $N_{i}$, of the grid is given proportionally to the total number of packet simulated, $N_{\rm tot}$, from the contribution of the cell, $L_{i}$ [erg s$^{-1}$], to the total luminosity of all cells together $L_{\rm tot}$ [erg s$^{-1}$]
\begin{equation}
N_{\rm ph,i} = N_{\rm tot}\frac{L_{i}}{L_{\rm tot}}.
\end{equation}
However, for cells that contribute negligibly to the total luminosity (L$_{i}$ $\ll$ L$_{\rm tot}$), very few packets are simulated originating from the cell, leading to increased noise and not properly accounting for their contribution to the end spectrum.
To counter this, we apply the hybrid biasing scheme from \citet{Baes2016}, which combines a uniform in cell biased function with the non-biased function.
We have found high biasing is generally required to adequately capture the contribution from cooler regions of the planetary atmosphere, and is key in producing accurate phase curve spectra for the large temperature contrasts present in hot Jupiter atmospheres.

\subsection{Survival biasing}
In the pure MCRT set-up, after the packet has traveled a distance according to the sampled optical depth, a random number, $\zeta$, is drawn and compared to the local single scattering albedo, $\omega_{i}$, of the cell.
Should $\zeta$ $<$ $\omega_{i}$, the packet is scattered according to a given phase function, otherwise it is absorbed by the medium and its evolution discontinued.
Survival biasing instead forces the packet to scatter, but reduces the weight of the packet by the single scattering albedo, ensuring that energy is conserved in the simulation while allowing the packet to continue its journey.

\subsection{Russian roulette}
As a consequence of several of the above biasing schemes, depending on the simulation set-up, the photon packet can scatter in the medium indefinitely with ever decreasing weight.
In order to stop simulating the packet's path, a Russian roulette scheme is used \citep{Dupree2002}, whereby after the weight of a packet becomes below a threshold, it has a chance to be removed from the simulation.

\subsection{Next event estimation}
\label{sec:peeloff}
The next event estimation method, also known as the peeloff method, combines the regular MCRT model with geometric raytracing applied at every scattering or emitting location of the packet \citep{Yusef-Zadeh1984, Wood1999}.
This greatly reduces sampling noise and is the main tool that provides high fidelity output for the simulation.
The fraction, $f$, of a packet's luminosity escaping towards the chosen observational direction during a scattering event is given as

\begin{equation}
f = W_{ph}\omega \Theta(\theta_{obs}) \exp(-\tau_{obs}),
\end{equation}
where $\Theta(\theta_{obs})$ is the normalised scattering phase function towards the observational direction, $\omega$ the single scattering albedo, and $\tau_{obs}$ the total extinction optical depth towards the observational direction.

For emission spectra, should multiple-scattering not be required an initial peeloff from the emission location is enough to produce the spectrum.
To increase the efficiency of the calculation, a limiting vertical optical depth can be set to avoid simulating packets from very high optical depths that contribute negligibly to the output spectrum (here typically $\tau_{\rm vert}$ = 30).
This focuses the computational effort to produce low noise results from the emission in important photospheric regions.

\subsection{G-ordinance biasing}
When modelling using correlated-k opacities, the g-ordinance is required to be randomly sampled from the cumulative weight distribution of the g-ordinances \citep{Lee2019}.
For an accurate result, the opacity distribution should be adequately sampled across the range of weights.
However, for regions of low packet counts, this distribution may not be as evenly sampled as desired.

For unbiased g-ordinance sampling the probability, $p(g)$,  is given directly by the weight of the g-ordinance, $w_{g}$,
\begin{equation}
p(g) = w_{g}.
\end{equation}
For direct sampling of the k-distribution weights, we can construct a simple biased PDF, $q(g)$, to equally sample all g-ordinances
\begin{equation}
\label{eq:uniform}
q(g) = \frac{1}{N_{g}},
\end{equation}
where $N_{g}$ is the total number of g-ordinance points, with the weighting factor
\begin{equation}
W_{ph}(g) = \frac{p(g)}{q(g)} = N_{g}w_{g}.
\end{equation}

For sampling g-ordinances in emission, the unbiased g-ordinance sampling is given by
\begin{equation}
p(g) = \frac{w_{g}L_{g}}{\sum_{g}w_{g}L_{g}},
\end{equation}
where $L_{g}$ [erg s$^{-1}$] is the luminosity of that g-ordinance.
For the biased sampling, the same uniform probability as Eq. \ref{eq:uniform} can be assumed.
The weighting factor is then
\begin{equation}
W_{ph}(g) = \frac{p(g)}{q(g)} = \frac{N_{g}w_{g}L_{g}}{\sum_{g}w_{g}L_{g}}.
\end{equation}

We have also included a composite biasing method for sampling the g-ordinances.
In testing we have found this biasing to be vital when using k-tables that use split quadrature zones, for example, the 8+8 points used petitRADTRANS \citep{Molliere2019} and the 4+4 points used in SPARC/MITgcm \citep{Kataria2013}.
Due to the low weights for the high split quadrature values, the non-biased scheme will rarely sample these g-ordinance values, resulting in too low an opacity distribution being sampled.
The biasing scheme fixes this issue in a simple manner and ensures a more even sampling of the opacity distribution.
We have found that uniform sampled g-ordinance values perform similarly with or without this biasing.
However, we recommend this biasing for all cases anyway to ensure that the low weighted g-ordinances are adequately sampled.

\section{Atmospheric observables from gCMCRT}
\label{sec:obs}

In this section, we detail how albedo, transmission and emission spectra are calculated from the gCMCRT output.

\subsection{Transmission spectra}

gCMCRT has the ability to calculate fully 3D transmission spectra including the effects of multiple scattering.
The transmission spectra equation is given by \citep[e.g.][]{Dobbs-Dixon2013,Robinson2017}
\begin{equation}
\label{eq:trans}
\left(\frac{R_{p, \lambda}}{R_{\star}}\right)^{2} = \frac{1}{R_{\star}^{2}} \left(R_{p, 0}^{2} + 2\int_{R_{p, 0}}^{\infty}[1 - \mathcal{T}(b)]bdb\right),
\end{equation}
where R$_{p, \lambda}$ [m] is the wavelength dependent radius of the planet, R$_{\star}$ [m] the radius of the host star, R$_{p, 0}$ [m] the bulk planetary disk radius, $\mathcal{T}$ the transmission function, and $b$ [m] the impact parameter.
Formally the upper limit for the integral in Eq. \ref{eq:trans} is $\infty$.
This is replaced by the top of atmosphere radius, R$_{p, TOA}$ [m], as per the simulation boundaries to facilitate numerical calculations.

Following the principles of integration through independent sampling, the result of the integral in Eq. \ref{eq:trans}, I$_{p}$, is approximated by simulating a suitably large number of N$_{ph}$ photon packets that sample the integral function
\begin{equation}
\langle I_{p}\rangle = \frac{(R_{p, TOA} - R_{p, 0})}{N_{ph}} \sum^{N_{ph}}_{i}W_{\rm ph}[1 - e^{-\tau_{i}}]b_{i},
\end{equation}
where now each packet contributes its transmission for a randomly sampled impact parameter.
This method avoids the slight geometric biasing found in the original method in \citet{Lee2019}, where packets were binned in impact parameter.

An initial peeloff (raytracing step) is performed at the packet's initial transmission location, giving a baseline signal for the absorption spectra.
The packet is then free to multiply scatter through the atmosphere, each time contributing to the transmission through the atmosphere through the next event estimation method.
This weights the transmission with the scattering probability towards the observation direction, so packets can contribute to the transmission at different impact parameters from their initial position.
Packets are given a random starting location on the transmission annulus of the simulation, ensuring that raytracing from the packet's ensuing transit chord will most likely not occult the planetary surface boundary which would not contribute to the signal.

Should multiple scattering not be required, it can be switched off which reduces the model to a raytracer scheme with random transit chord sampler.
This significantly increases the efficiency of the model and is useful for when it is known multiple scattering will not be an important contribution to the transmission spectrum, which is typically the case for infrared wavelength regimes \citep[e.g.][]{Hubbard2001}.

\subsubsection{Limb darkening}

An important consideration for transmission spectra calculations is the change in the stellar disk intensity that passes through the atmosphere as the planet transits across the stellar disk.
For our transmission spectra modes we include the capability to take the limb darkening of the star into account.

To calculate this, the sampled transmission chord is mapped using spherical geometry directly to a longitude and latitude, and therefore a zenith angle, $\mu$, onto the spherical stellar surface.
This cosine angle is then converted to the limb darkening fraction, $I_{\rm ph}(\mu)/I(0)$, the ratio of the intensity at $\mu$ to the zenith point, using a limb darkening law, for example, the quadratic law \citep{Kopal1950}
\begin{equation}
\frac{I_{\rm ph}(\mu)}{I(0)} = 1 - c_{1}(1 - \mu) - c_{2}(1 - \mu)^{2},
\end{equation}
where $c_{1}$ and $c_{2}$ are fitting coefficients.
We include all limb darkening laws detailed in J. Southworth's website\footnote{https://www.astro.keele.ac.uk/jkt/codes/jktld.html} with user supplied limb darkening coefficients.

Each packet then has a unique limb darkening fraction value, which is then used to reduce the weight of the packet through
\begin{equation}
W_{ph, new} = W_{ph} \frac{I_{ph}(\mu)}{I(0)},
\end{equation}
which is then the correctly weighted packet's contribution to the transmission signal.

\subsection{Scattered incident light}
The geometric albedo, A$_{g}$ is defined as the ratio of incident to reflected flux at zero phase \citep[e.g.][]{Marley1999,Heng2021}
\begin{equation}
A_{g} = \frac{j(0)}{\pi F_{\rm inc}},
\end{equation}
where j(0) is the outgoing flux at 0 phase and F$_{\rm inc}$ the incident flux.
In gCMCRT, the reflected fraction of the energy escaping towards the observable direction is directly calculated, with the incident flux normalised to 1.
For a geometric albedo calculation therefore, the results for a reflection calculation at 0 phase is simply multiplied by $\pi$ to account for the definition of A$_{g}$.
The geometric albedo can also be given as a function of wavelength.

For scattered light phase curves, the viewing angle of the detector can be changed for the desired phases.
The output is then scaled to the reflected fraction at zero phase, essentially deriving the normalised classical planetary phase function for the simulated planet, given by
\begin{equation}
\Phi(\phi) = \frac{j(\phi)}{j(0)}.
\end{equation}
The planet-to-star flux ratio is then given by the well known formula
\begin{equation}
\frac{F_{\rm p}}{F_{\star}}(\phi) = \left(\frac{R_{\rm p}}{a}\right)^{2}A_{g}\Phi(\phi),
\end{equation}
where R$_{\rm p}$ [cm] is the planetary radius and $a$ [cm] the semi-major axis.

It is useful to note that in this case $\Phi(\phi)$ can be $>$ 1 dependent on the variety of scattering components in the atmosphere, for example a bright westward cloud, allowing non-symmetrical phase curves to be modelled.


\subsection{Emitted planetary light}

To calculate emission spectra each photon packet is emitted from within the atmospheric volume and its contribution to the outgoing energy towards a certain viewing angle is tracked.
For cloud free simulations, scattering is generally negligible and can be switched off for faster results, especially when focusing on infrared wavelengths.

The luminosity of an individual cell i, L$_{\lambda}$ [erg s$^{-1}$ cm$^{-1}$], for a given wavelength $\lambda$ is given by
\begin{equation}
L_{\lambda, i} = 4\pi V_{i} \rho_{i} \kappa_{\rm abs, i} B_{\lambda}(T_{i}),
\end{equation}
where $V_{i}$ [cm$^{3}$] is the volume of the cell, $\rho_{i}$ [g cm$^{-3}$] the density of the cell, $\kappa_{\rm abs, i}$ [cm$^{2}$g$^{-1}$] the absorption opacity of the cell (including any cloud absorption opacity) and $B_{\lambda}$(T$_{i}$)[erg s$^{-1}$ cm$^{-2}$ cm$^{-1}$ sr$^{-1}$] the Planck function at temperature T$_{i}$.

The luminosity escaping towards the observable direction, L$_{\rm p}$ [erg s$^{-1}$ cm$^{-1}$] , is then
\begin{equation}
L_{\rm p} = \frac{L_{tot}}{N_{ph}}\sum_{i}\sum^{N_{ph}}_{n}  f(\phi,\theta),
\end{equation}
where L$_{tot}$ = $\sum$ L$_{\lambda}$.

The specific intensity emanating from the planet toward a certain viewing angle, I$_{p}$ [erg s$^{-1}$ cm$^{-2}$ cm$^{-1}$] is then given by
\begin{equation}
I_{\rm p} = \frac{L_{\rm p}}{\pi (R_{\rm p, 0} + z_{\rm TOA})^{2}},
\end{equation}
where $z_{\rm TOA}$ [cm] is the top of atmosphere altitude.
In this case the denominator of this equation is always an area of a circle, as the output of gCMCRT is the luminosity escaping from a spherical hemisphere of the planet.
The flux from the planet, F$_{\rm p}$ [erg s$^{-1}$ cm$^{-2}$ cm$^{-1}$], is then given by
\begin{equation}
F_{\rm p} = \pi I_{\rm p},
\end{equation}
as per the definition of converting specific intensity to flux from a spherical surface.

The planet-to-star ratio $F_{\rm p}/{F_{\star}}$  is then given by using a stellar model atmosphere \citep[e.g.][]{Kurucz1995}, interpolated to the same wavelength grid of the simulation.
To produce thermal phase curves, the simulation is run on the GPU several times for different viewing angles and the results reconstructed after the simulation is complete.

By directly accounting for the volume of the emitting regions, gCMCRT avoids the so-called wavelength dependent `photospheric radius problem', present when modelling emission spectra from non-uniform T-p structures across a globe \citep[e.g.][]{Dobbs-Dixon2017, Fortney2019}.
gCMCRT therefore correctly weights the hotter and cooler parts of the atmosphere contribution to the end spectrum, avoiding known biases with blended hotter and cooler T-p profiles \citep[e.g.][]{Feng2016, Taylor2021}.

\section{Albedo and emission contribution functions}
\label{sec:confunc}

We include the ability in gCMCRT to compute 3D contribution functions.
The contribution function is generally given by \citep[e.g.][]{Drummond2018c}
\begin{equation}
\label{eq:CF}
CF = B_{\lambda}(T) \frac{d\exp(-\tau/\mu)}{d\log_{10}{P}},
\end{equation}
where the $\log_{10}$P term scales the contribution per decade of pressure level.
For the height vertical grid, spherical 3D geometry case, and with the random starting positions of MCRT packets, this pressure term is non-trivial to calculate.
Instead in gCMCRT, for albedo spectra, emission spectra and phase curves, the contribution function calculation performed by tracking the total contribution each cell made to the output spectra then normalising by the total escaped energy fraction
\begin{equation}
CF(\theta, \phi, R) = \frac{f(\theta, \phi, R)}{\sum f(\theta, \phi, R)}.
\end{equation}
A similar counter is used for the scattering component, therefore allowing the discretion between the thermal and scattering contributions to the result.
This contribution function is different to Eq. \ref{eq:CF}, as it directly gives the fractional contribution of that cell to the end spectra, and is not normalised to the column maximum contribution.
The same method can also be used to calculate the contribution functions for albedo spectra.

\section{Opacities}
\label{sec:opac}

\begin{table}
\centering
\caption{Line opacity species and references used for the publicly available gCMCRT formatted k-tables at R = 100 resolution.}
\begin{tabular}{c c}  \hline \hline
Species &  Reference \\ \hline
Na  &  \citet{Kurucz1995}   \\
K   &  \citet{Kurucz1995}  \\
Fe   &  \citet{Kurucz1995}  \\
Fe$^{+}$   &  \citet{Kurucz1995}  \\
H$_{2}$O   &  \citet{Polyansky2018}   \\
OH   &  \citet{Hargreaves2019}   \\
CH$_{4}$   &  \citet{Hargreaves2020}  \\
C$_{2}$H$_{2}$   &  \citet{Chubb2020}   \\
CO   &  \citet{Li2015}   \\
CO$_{2}$    &  \citet{Yurchenko2020}   \\
NH$_{3}$   &  \citet{Coles2019}   \\
HCN   & \citet{Harris2006, Barber2014} \\
H$_{2}$S   & \citet{Azzam2016} \\
SH & \citet{Gorman2019} \\
HF & \citet{Li2013,Coxon2015} \\
SiO & \citet{Yurchenko2021} \\
TiO & \citet{McKemmish2019} \\
VO & \citet{McKemmish2016} \\
FeH & \citet{Bernath2020} \\
\hline \hline
\end{tabular}
\label{tab:line-lists}
\end{table}

Accompanying gCMCRT we provide an opacity mixer and interpolator (optools) based on the \citet{Lee2019} model with some updates.
This calculator takes in the same flattened 1D array of temperature, pressures, mean molecular weight and gas species volume mixing ratios as gCMCRT and calculates the gas phase line opacities from pre-calculated cross section or k-coefficient tables.
Currently this code is parallelised using openMP and requires CPUs to operate, however, plans to overhaul optools for GPU computing are underway.

We create custom k-tables using cross sections calculated using the HELIOS-K \citep{Grimm2015,Grimm2021} opacity calculator as well as the EXOPLINES opacities \citep{Gharib-Nezhad2021} spanning YY wavelength bands over 0.3-30 $\mu$m at R = 100 resolution, suitable for producing JWST predictions.
These k-tables use an 8+8 g-ordinance scheme, with 8 g-ordinances between g = 0-0.9 and 8 between g = 0.9-1.0, the same as petitRADTRANS \citep{Molliere2019}.
The line-lists used to generate the cross sections for the k-tables are given in Table. \ref{tab:line-lists}.
The original cross section data were calculated at a resolution of 0.01 cm$^{-1}$.
NEMESIS \citep{Irwin2008} formatted k-tables available from ExoMol \citep{Chubb2021} can be also used directly within the opacity mixer, which were benchmarked extensively in \citet{Lee2019}.
We also include conversion scripts for the ARCiS \citep{Min2020} and petitRADTRANS \citep{Molliere2019,Molliere2020} k-tables and the TauREx \citep{Al-Refaie2021} cross-section tables, also available on ExoMol, to the gCMCRT format.
Individual species k-tables are mixed using the random overlap method \citep[e.g.][]{Lacis1991,Amundsen2017}.
This capability allows the user to take the same opacities commonly used in 1D modelling codes and apply them directly to the 3D model.

We also provide the ability to interpolate pre-mixed k-tables to the atmospheric T-p structure.
The GitHub repository contains a pre-mixed 1x solar metallicity opacity k-table, one with TiO and VO and one without, suitable for general post-processing use.
Equilibrium condensation is included in the table.
Other pre-mixed tables are available from the repository as well.

We take CIA opacities from the HITRAN database \citep{Karman2019}, and options to calculate H$^{-}$ bound-free and free-free opacities following \cite{John1988} and He$^{-}$ free-free opacity following \citet{Kurucz1970}.
For Rayleigh scattering opacities we include cross-section and refractive index data from various sources, namely \citet{Kurucz1970, Sneep2005, Irwin2009, Thalman2014} and \citet{Wagner2008} for H$_{2}$O.
Thomson cross-sections from free electrons can also be calculated as a Rayleigh scattering contribution.

For cloud opacities, single scattering albedos and asymmetry factors, we use the Mie calculator MieX \citep{Wolf2004}.
Various analytical cloud size particle distributions from the literature can also be calculated with optools (e.g. log-normal \citep{Ackerman2001} and potential exponential \citep{Helling2008}), in addition to numerical size distributions calculated by codes such as DRIFT \citep[e.g.][]{Helling2008}.
Bin based microphysical cloud models such as CARMA \citep{Powell2018,Gao2020} can also be processed using optools.
CIA, Rayleigh and cloud opacities are calculated at the bin center wavelengths in correlated-k mode.

For high resolution spectral modelling we include the option to include the Doppler shift of lines in the line of sight due to wind velocity and planetary rotation.
Taking the horizontal, meridional and vertical velocity (u, v, w [cm s$^{-1}$] respectively) input from the GCM model, the line of sight velocity, v$_{\rm LOS}$ [cm s$^{-1}$], is given by \citep[][App. B]{Wardenier2021}
\begin{eqnarray}
v_{\rm LOS} = u\sin(\theta_{v})\sin(\phi - \phi_{v}) \nonumber \\
+ \left[v\cos(\theta) - w\sin(\theta)\right]\sin(\theta_{v})\cos(\phi - \phi_{v}) \nonumber\\
- \left[v\sin(\theta) + w\cos(\theta)\right]\cos(\theta_{v}) \nonumber \\
+ \Omega\left(R_{0} + z\right)\sin(\theta)\sin(\theta_{v})\sin(\phi - \phi_{v}),
\end{eqnarray}
where $\Omega$ [rad s$^{-1}$] is the rotation rate of the planet, R$_{0}$ [cm] the lower boundary planetary radius, z [cm] the cell altitude, $\theta$ the latitude of the cell and $\phi$ the longitude of the cell.
$\theta_{v}$ and $\phi_{v}$ are the viewing latitude and longitude respectively.
The opacity in each cell is then interpolated to the Doppler shifted effective wavelength, $\lambda_{\rm eff}$ [$\mu$m],
\begin{equation}
\lambda_{\rm eff} = \lambda_{0} \left(1 - \frac{v_{\rm LOS}}{c}\right),
\end{equation}
where $\lambda_{0}$ [$\mu$m] is the rest frame wavelength and $c$ [cm s$^{-1}$] the speed of light.

The Doppler shifting of the opacities in each cell is performed at runtime inside gCMCRT.
We employ a moving wavelength window method, where, rather than read in all opacity data across the required wavelength range to interpolate to, opacity data across a positive and negative wavelength range from the rest wavelength are retained in memory and Doppler shifted opacity interpolated to this window.
The window is then moved one wavelength forward for the next wavelength iteration, retaining the rest wavelength in the central part of the block.
This avoids the large memory requirements when storing large amounts of opacity data across the 3D GCM grid.
The size of this moving window can be changed to optimise the calculation, depending on how shifted the opacities are in the line of sight.

\section{Gas phase abundances}
\label{sec:CEab}

gCMCRT contains no native way to produce gas phase abundances from GCM model output.
However, we include 2D T-p grids of gas phase volume mixing ratios and mean molecular weights in chemical equilibirum calculated using the ggChem \citep{Woitke2018} code.
These chemical tables can then be interpolated to the GCM T-p profiles using a python script.
This grid is what is used in the current study to find gas phase abundances when the GCM does not contain chemical information.
Alternatively, chemical information can be converted directly from the GCM or other chemical scheme to the gCMCRT profile format.

\section{Albedo spectra}
\label{sec:albspec}

\begin{figure} 
   \centering
   \includegraphics[width=0.49\textwidth]{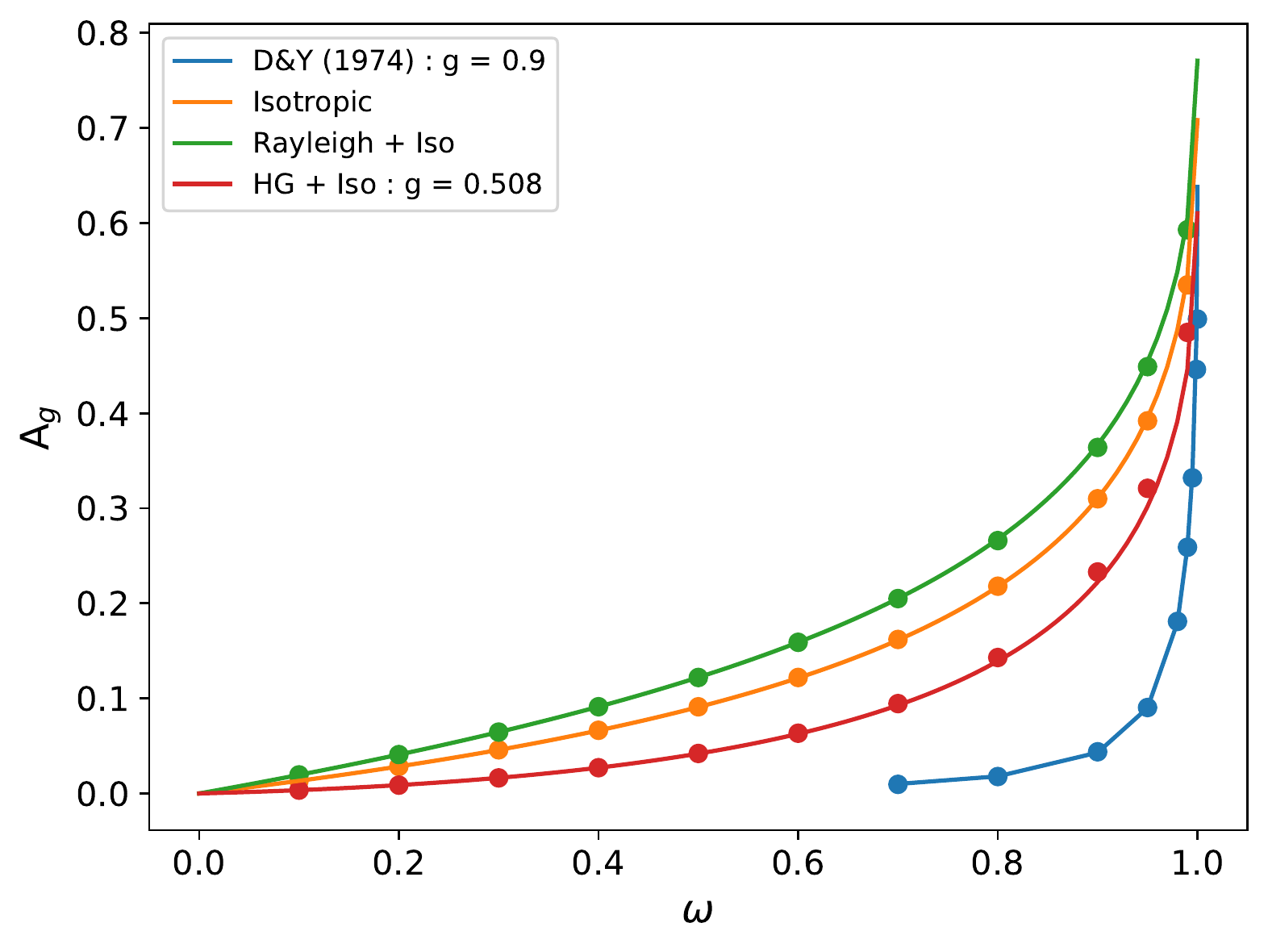}
   \caption{Geometric albedo, A$_{\rm g}$ comparison to analytical and known solutions for various phase functions.
   The dots show the solutions from gCMCRT and the solid lines the analytical/known solutions.
   From \citet{Heng2021}: orange - isotropic multiple scattering solution, green - Rayleigh scattering with isotropic multiple scattering, red - HG scattering with isotropic multiple scattering for $g$ = 0.508.
   The comparison to the \citet{Dulgach1974} values (HG multiple scattering) for $g$ = 0.9 are shown in the blue colour.}
   \label{fig:Ab}
\end{figure}

\begin{figure*} 
   \centering
   \includegraphics[width=0.49\textwidth]{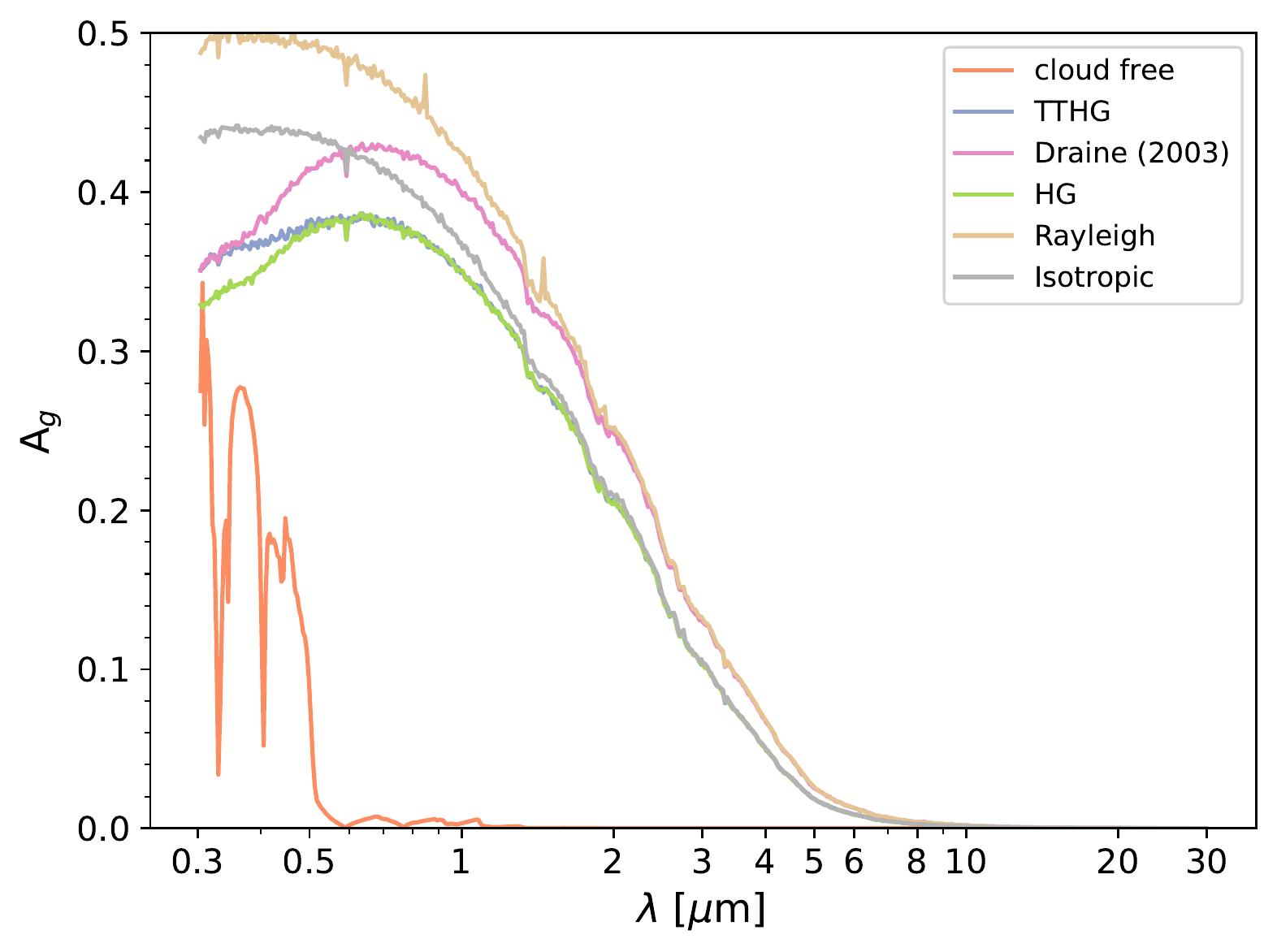}
   \includegraphics[width=0.49\textwidth]{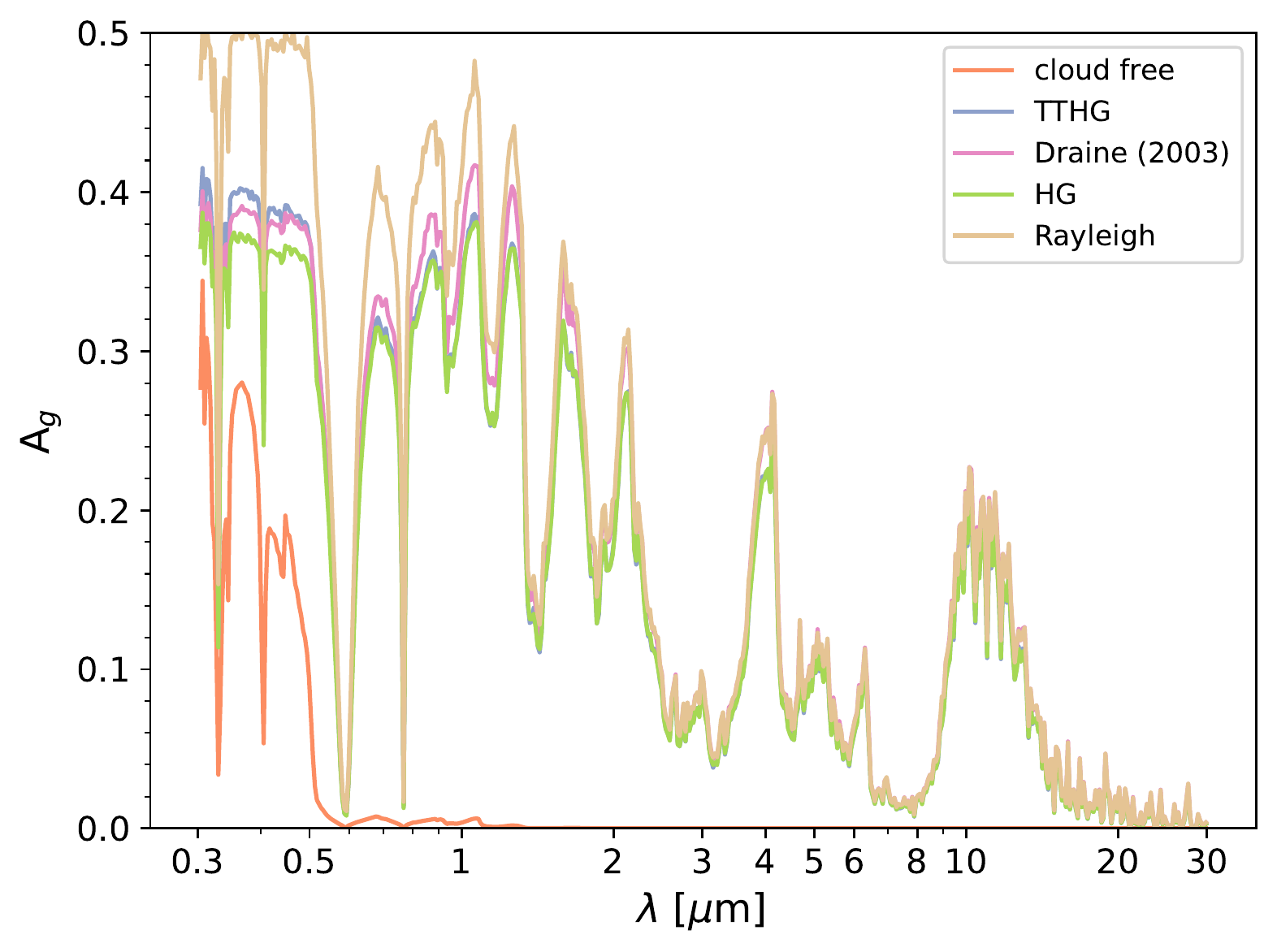}
   \caption{Geometric albedo calculations for the T$_{\rm eq}$ = 2250 K, g = 10 m s$^{-1}$ \citet{Roman2021} cloudy GCM models.
   Left: Nucleation limited, extended clouds scenario.
   Right: Nucleation limited, compact clouds scenario.
   Coloured lines show the albedo assuming different scattering phase functions for the cloud particles.}
   \label{fig:Roman}
\end{figure*}

We first check the algorithm for the albedo spectra mode by comparing the numerical results of gCMCRT to known analytical solutions.
In Figure \ref{fig:Ab} we compare the geometric albedo to the analytical solutions in \citet{Heng2021}, who presented scattering functions for certain phase functions with isotropic multiple scattering.
We also compare to the \citet{Dulgach1974} Henyey-Greenstein solutions at high $g$ ($g$ = 0.9).
Our model is able to reproduce the analytical solutions well, with only larger deviations occurring at very high single scattering albedo ($\omega$  $>$ 0.99).

For a 3D in-situ demonstration we use GCM output from the \citet{Roman2021} grid of cloudy models.
We post-process two models, the T$_{\rm eq}$ = 2250 K, g = 10 m s$^{-1}$ nucleation limited model with extended clouds and the same model but with compact clouds \citep{Roman2021}.
We assume the effective particle size given by the vertical size dependent scheme in \citet{Roman2021}.
For the cloud opacity calculation we assume a log-normal distribution with variance 0.1 $\mu$m as in \citet{Roman2021}.

Figure \ref{fig:Roman} shows the results of post-processing the two GCM simulations.
It is clear that the extended clouds and compact clouds produce dramatically different results.
The extended clouds show a Mie scattering slope from the optically thick high altitude clouds, of particle size $\sim$ 0.1 $\mu$m.
This particle size has a maximum scattering efficiency at approximately 0.6 $\mu$m, the size parameter of 0.1 $\mu$m particles.
The compact clouds produce a more wavelength dependent albedo spectra, due the efficiency of scattering from larger particle sizes deeper in the atmosphere.
The spectra shows gas extinction features where the gas is more opaque, given by the packets scattered off the compact, deep cloud layer and escaping back through the gas towards the observer.

In Fig. \ref{fig:Roman} we also produce the effect of different scattering phase functions for the cloud particles.
We perform the MCRT with five phase function tests, one cloud free (without cloud opacity), one with the TTHG function, the HG function, Rayleigh function, the \citet{Draine2003} function and an isotropic function for the scattering from cloud particles.
From the results, the calculated geometric albedo can vary significantly depending on the chosen cloud particle scattering phase function.
The Rayleigh function provides the largest albedo due to the equal forward and backscattering fractions.
The TTHG and HG function are generally similar, except at optical wavelengths where the backscattering term in the TTHG becomes more important.
The \citet{Draine2003} function performs as expected, with more HG-like behaviour at optical wavelengths (and higher $g$), which returns to the Rayleigh scattering function as $g$ decreases at IR wavelengths.

\section{Emission spectra}
\label{sec:emspec}

\begin{figure*} 
   \centering
   \includegraphics[width=0.49\textwidth]{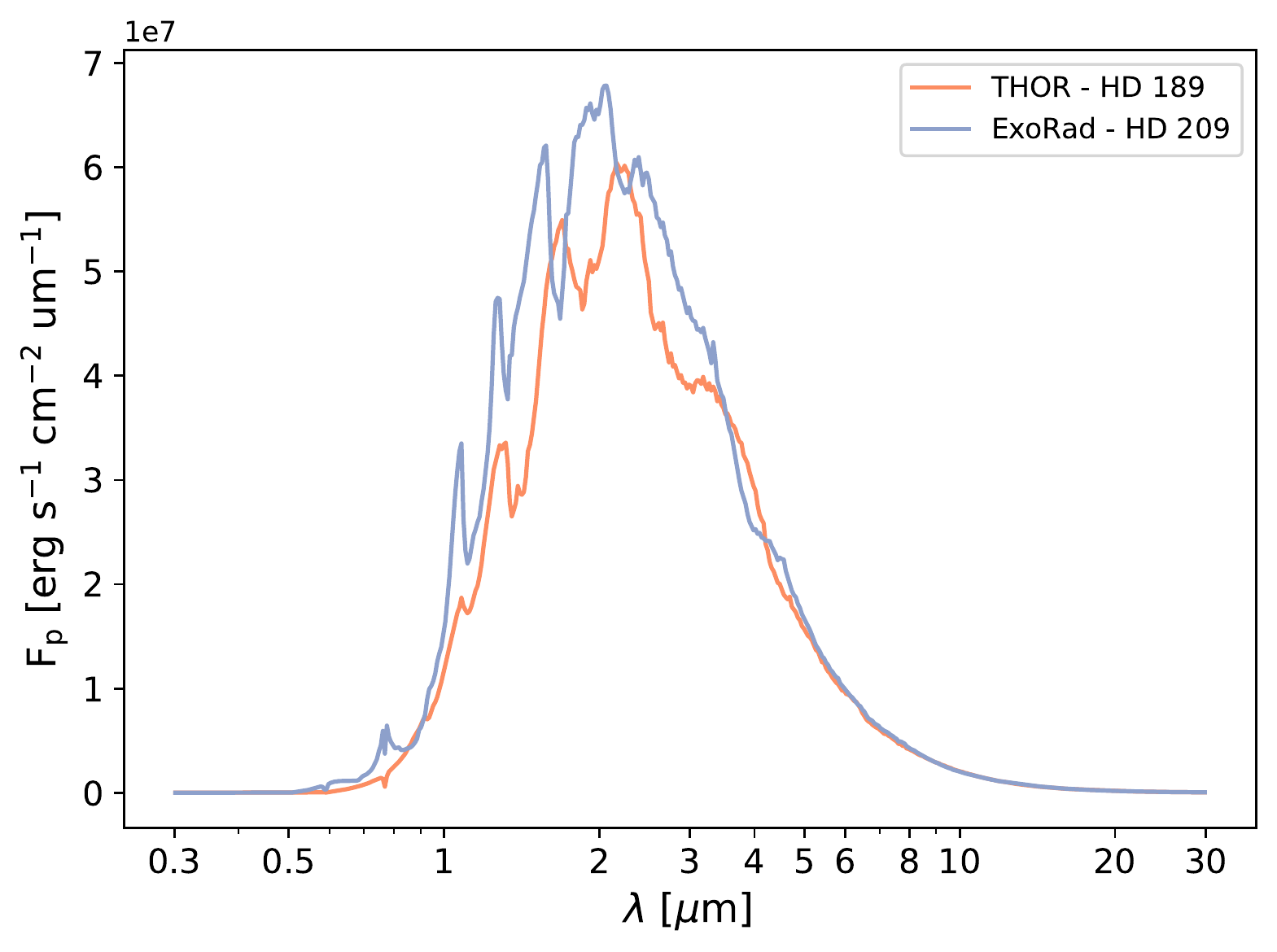}
   \includegraphics[width=0.49\textwidth]{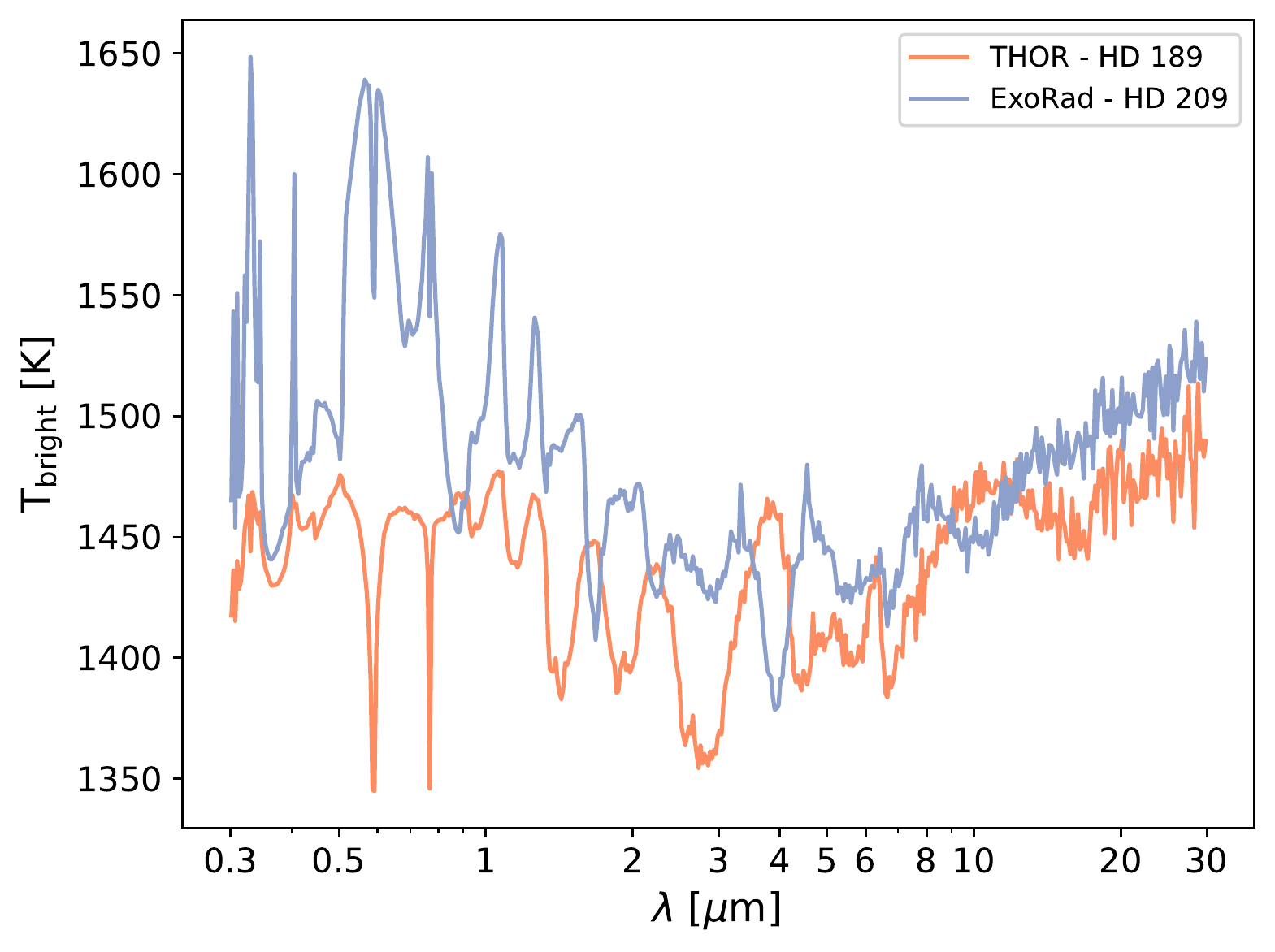}
   \caption{Left: Dayside emission spectra for the THOR and ExoRad models of HD 189733b and HD 209458b respectively.
   Right: Brightness temperature of the GCM models derived from the emission flux.}
   \label{fig:TM1}
\end{figure*}

\begin{figure*} 
   \centering
   \includegraphics[width=0.49\textwidth]{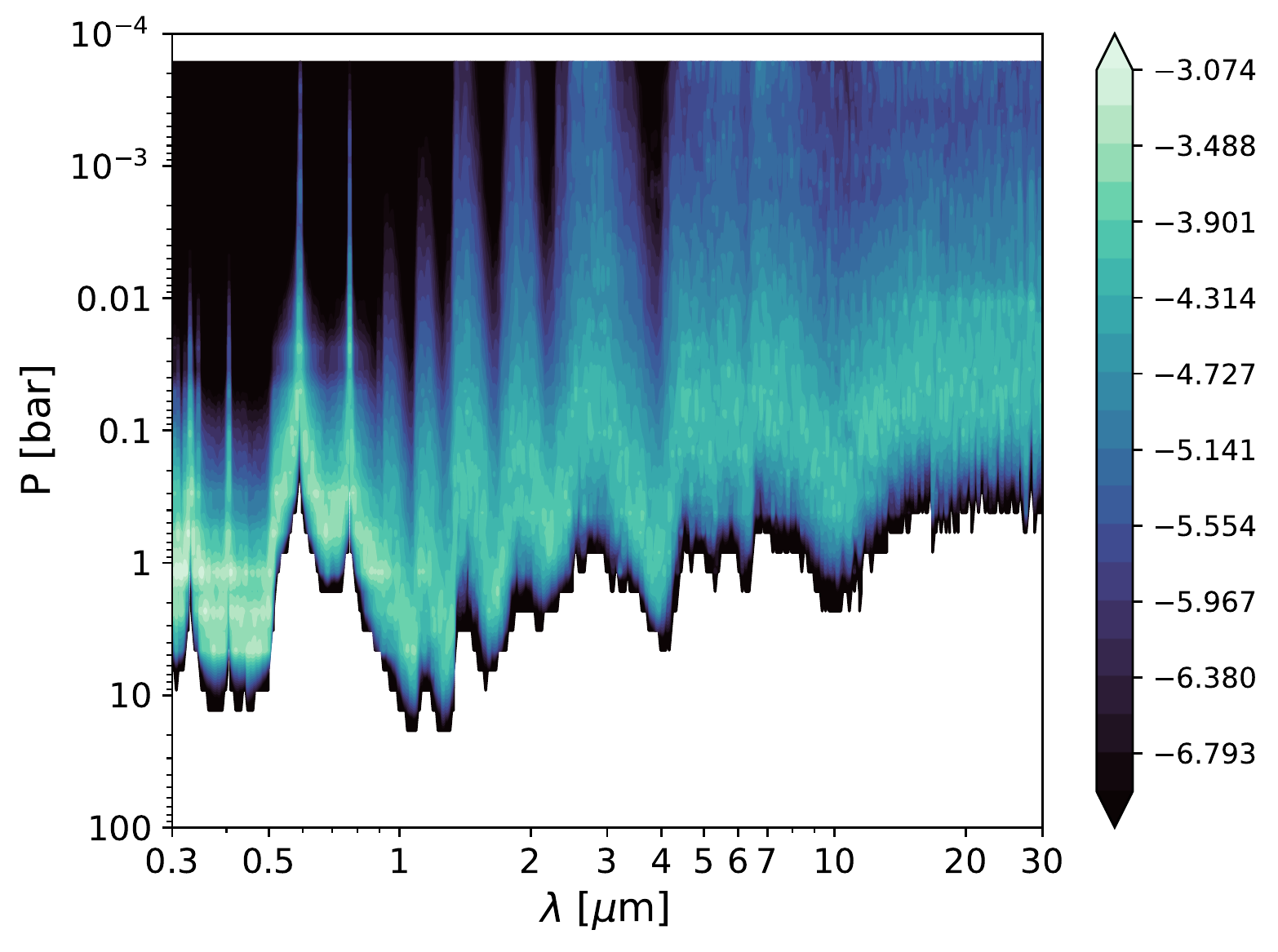}
   \includegraphics[width=0.49\textwidth]{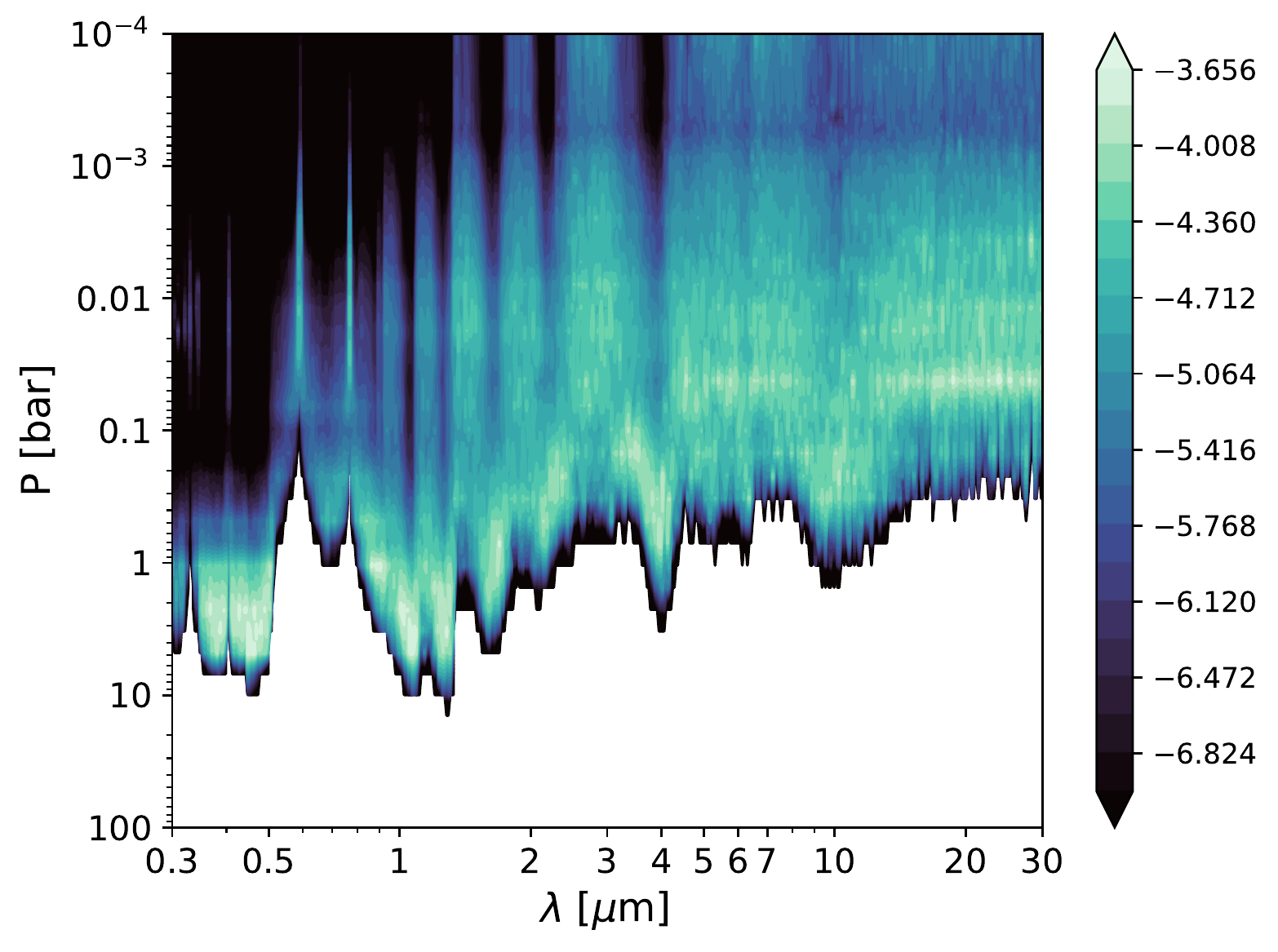}
   \caption{Contribution functions at the sub-stellar point.
   The colourbar shows the $\log_{10}$ of the contribution function.
   Left: THOR HD 189733b GCM model.
   Right: ExoRad HD 209458b GCM model.}
   \label{fig:TM2}
\end{figure*}

For emission spectra demonstration we post-process output from an HD 189733b-like THOR model \citep{Deitrick2020} and an HD 209458b-like simulation from the grid of \citet{Baeyens2021}, using the \citet{Carone2020} model (ExoRad).
In Fig. \ref{fig:TM1} we show the results of the dayside emission spectra of the two models.
Different absorption features can be distinguished in each model. Furthermore, the HD 189733b spectrum is generally lower, as expected given the lower equilibrium temperature of this planet.
In Fig. \ref{fig:TM2} we show the contribution functions of the two models at the sub-stellar point.
These plots show the fraction that the atmospheric column at the sub-stellar point contributed directly to the end spectrum, which is the contribution from all columns.
These show similar contribution functions for each model, showing the flux emanates from similar pressures at the sub-stellar point.

\section{Transmission spectra}
\label{sec:transspec}

\begin{figure} 
   \centering
   \includegraphics[width=0.49\textwidth]{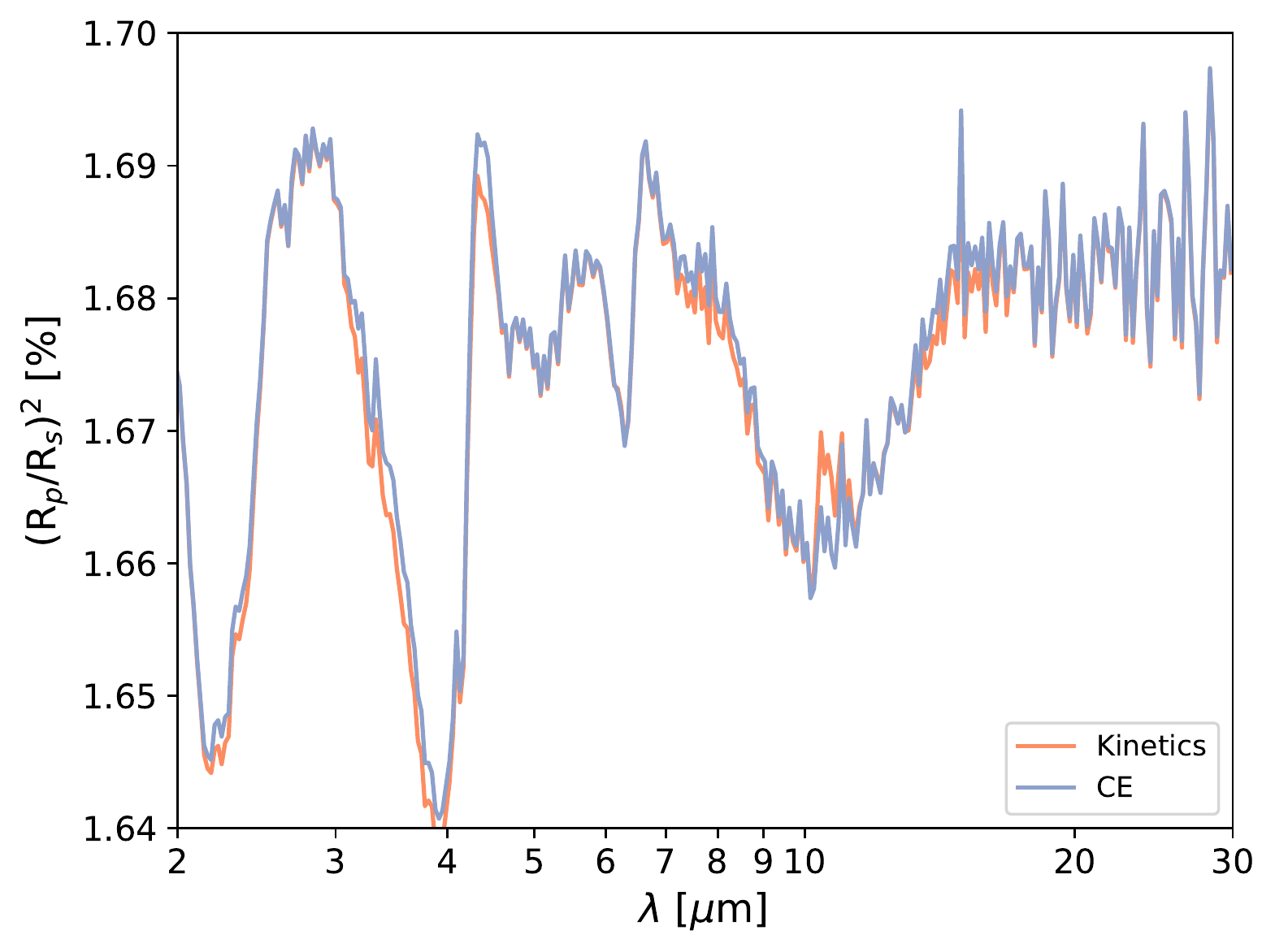}
   \caption{Modelled transmission spectra of HD 209458b from gCMRCRT using output from UK Met Office UM GCM \citep{Drummond2020}.
   Curves show spectra assuming from models assuming chemical equilibrium (blue) and non-equilibrium kinetic solutions (orange) results.
}
   \label{fig:UM}
\end{figure}

In this section we perform post-processing of the HD 209458b GCM models presented in \citet{Drummond2020}.
\citet{Drummond2020} investigated the influence of non-equilibrium chemistry by performing two simulations, one assuming chemical equilibrium (CE) and one with a kinetic chemical network.

We post-process both simulations using directly the chemical abundances from the GCM output.
In Fig. \ref{fig:UM} we show the transmission spectrum of the CE and kinetic chemistry models performed in \citet{Drummond2020}.
We are able to directly reproduce the findings in \citet{Drummond2020}, recreating the differences in the CE and kinetic models near 4.5$\mu$m due to the reduction of CO$_{2}$ in the kinetics scheme \citep{Drummond2020} and near 10$\mu$m due to an increase in the NH$_{3}$ abundance.

\section{High-resolution emission spectra}
\label{sec:hiem}

\begin{figure*} 
   \centering
   \includegraphics[width=0.49\textwidth]{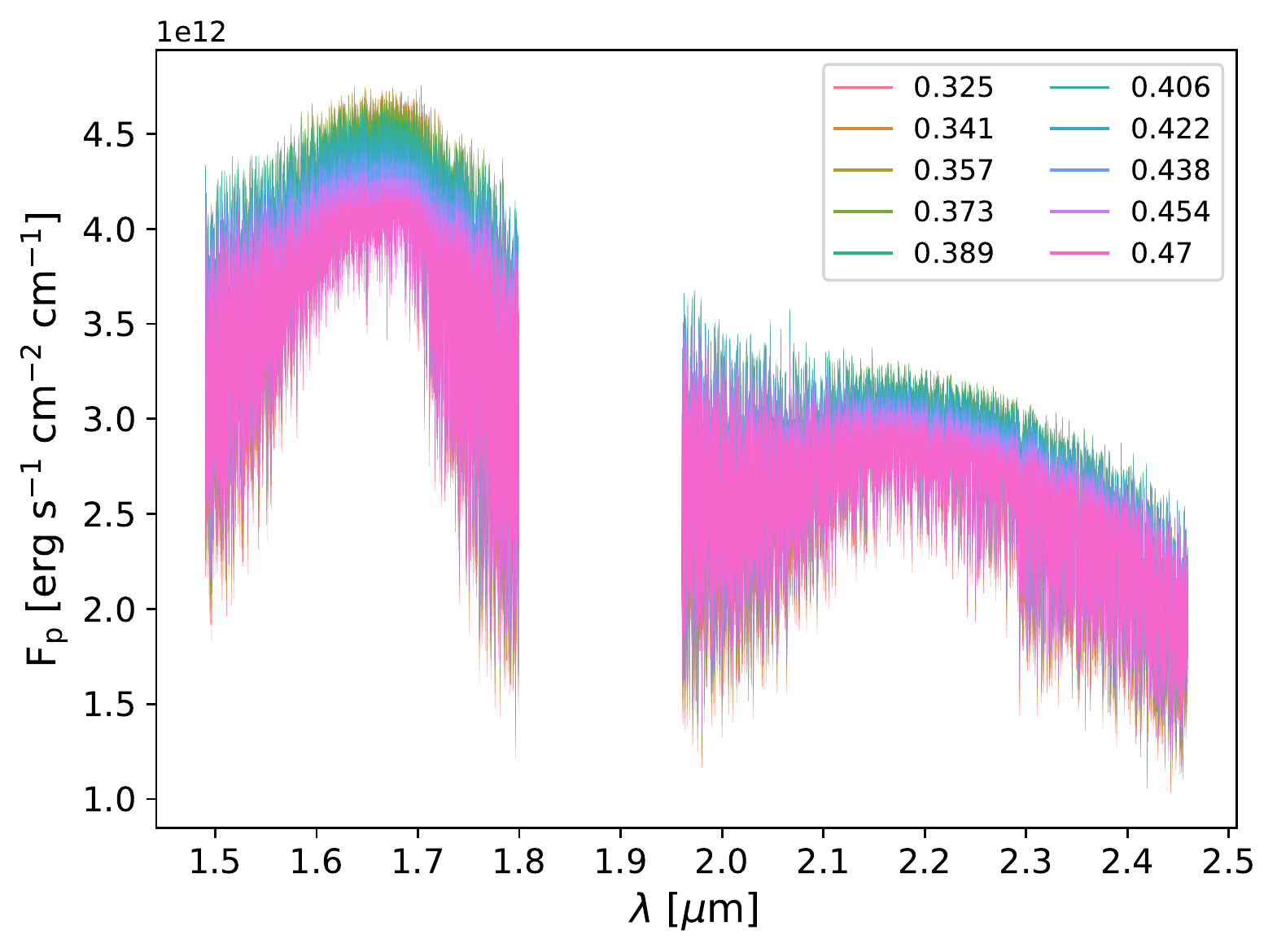}
   \includegraphics[width=0.49\textwidth]{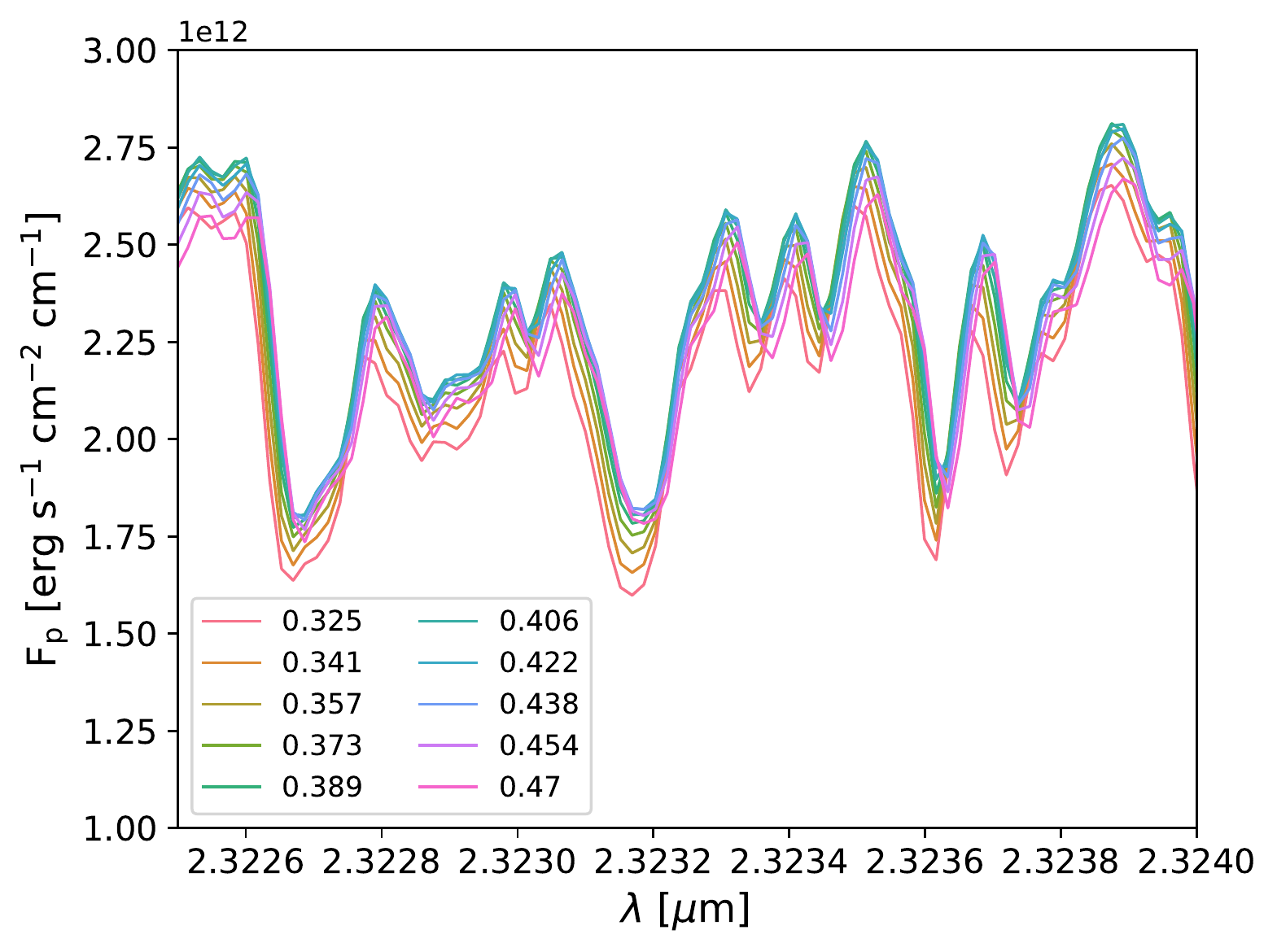}
   \caption{Post-processing emission spectra of the SPARC/MITgcm WASP-77Ab GCM at high-resolution in the H and K bands.
   Colours denote the phase of the planet.
   Left: Full H and K band results.
   Right: Zoomed portion of the emission spectra between 2.3225 and 2.3240 $\mu$m}
   \label{fig:W771}
\end{figure*}

\begin{figure*} 
   \centering
   \includegraphics[width=0.98\textwidth]{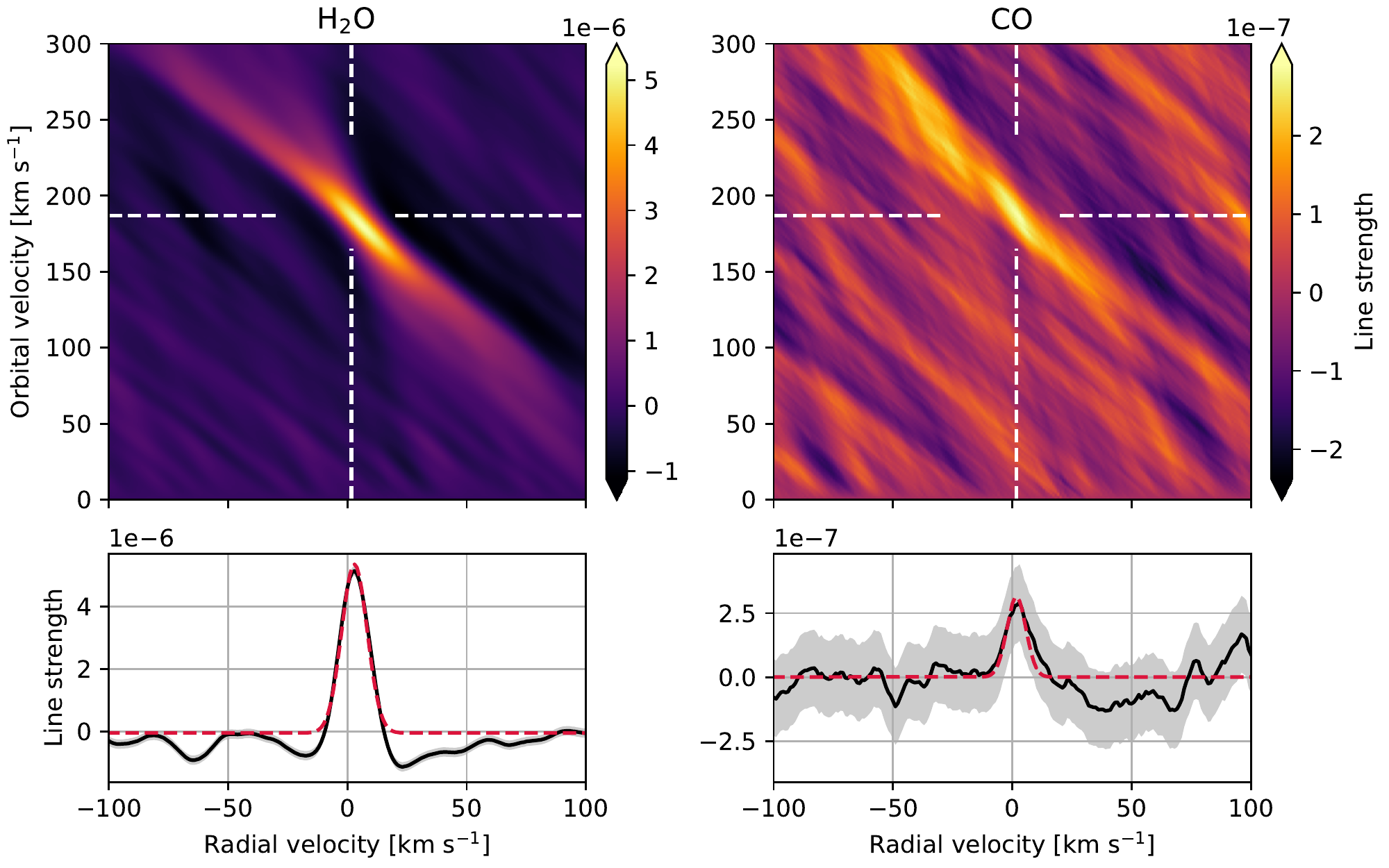}
   \caption{Upper panels: Velocity-velocity diagrams showing the modelled emission signals of H$_{2}$O and CO in the rest frame of the star.
   The horizontal dashed lines indicate the orbital velocity at which the signals were extracted ($v_{\rm orb} =  \frac{2 \pi a}{P} = 186.78\ {\rm km\ s}^{-1} $, see \citet{Cortes2020} for orbital parameters).
   The vertical dashed lines indicate the systemic velocity of $v_{\rm sys} = 1.68\ {\rm km\ s}^{-1}$ \citep{Line2021}. Lower panels: Cross-correlation functions stacked in the rest frame of the planet at the extracted orbital velocity.
   The shaded regions indicate the 1$\sigma$ uncertainties.
   Dashed lines show the best fit Gaussian model using only every forth point ($\Delta v_{\rm RV} = \ {\rm km\ s}^{-1} $) to account for correlation between the radial velocity points.}
   \label{fig:em_cc}
\end{figure*}

As a demonstration of the high-resolution phase curves we post-process a WASP-77Ab GCM simulation that was performed using the SPARC/MITgcm model \citep{Showman2009,Parmentier2021}.
We chose the IGRINS instrument \citep{Levine2018} H (1.45-1.8 $\mu$m) and K (1.96-2.45 $\mu$m) bands which have a native resolution of R = 45,000.
We therefore post-process the simulation at 3x the native resolution at R = 135,000, which was found to be high enough for modelling the cross-correlation signal, with 25517 wavelength points in the H band and 30675 wavelength points in the K band.
We model the planet following the observational strategy of \citet{Line2021}, who observed WASP-77Ab with IRGINS in the pre-eclipse phases between 0.325-0.47.

Figure \ref{fig:W771} shows the post-processing at high resolution of the full spectrum for the pre-eclipse phases and a zoomed in portion between 2.3225-2.3240 $\mu$m.
From this figure the gross behaviour of the simulated atmosphere can be seen, where both the shifting of the flux of the planet with phase and increases and decreases in the absolute flux.

We perform a synthetic cross-correlation on our modelled emission spectra phase curves, interpolating the modelled phases to a phase-grid corresponding to 500 second exposures during the orbit time series with $0.31 < \varphi < 0.47$ of the planet.
We produced H$_{2}$O and CO spectral templates by computing emission spectra at 0.5 phase, each with only H2O + continuum and CO + continuum opacity sources.
We then produced a high-resolution stellar spectrum for WASP-77Ab using SME \citep[][, version 580, private communication]{Piskunov2017}, which was then broadened to a projected rotational velocity of $v\sin{i}$ = 4.0 $\pm$ 0.2 km s$^{-1}$ \citep{Bonomo2017}, corrected for the Keplerian (K = 0.3234 km s$^{-1}$, \citep{Cortes2020}) as well as the systemic velocity (1.6845 $\pm$ 0.0004, \citep{Maxted2013}), resulting in spectra in the barycentre rest frame.
The modelled emission spectra were then shifted to the same rest-frame for each exposure by correcting for the planetary motion and the systemic velocity, and added to the high-resolution stellar spectrum.
Both spectra were then interpolated onto the wavelength grid of IGRINS, and noise with SNR = 200, assuming minimal SNR as found by \cite{Line2021}, and modelled tellurics were added.
The spectra were then split into the 54 IGRINS orders and the cross-correlation performed using the techniques following \citet{Hoeijmakers2019, Hoeijmakers2020, Prinoth2022}.
Figure \ref{fig:em_cc} shows the results of the emission signals, showing a detection of both H$_{2}$O and CO in our modelled planetary emission spectra.
We report a slight redshift of a few km s$^{-1}$ in both molecules from the GCM model results.
This is to be expected physically, since in transmission spectra the spectra is blushifted due to the equatorial jets being weaker on the western limb compared to the eastern limb.
When viewed from the dayside hemisphere, a redshift in therefore expected.
\citet{Line2021} report a blueshift in their modelling results, however, this may be a result of using 1D models in their analysis (M. Line per. comm.).

\section{High-resolution transmission spectra}
\label{sec:hitrans}

\begin{figure*} 
   \centering
   \includegraphics[width=0.49\textwidth]{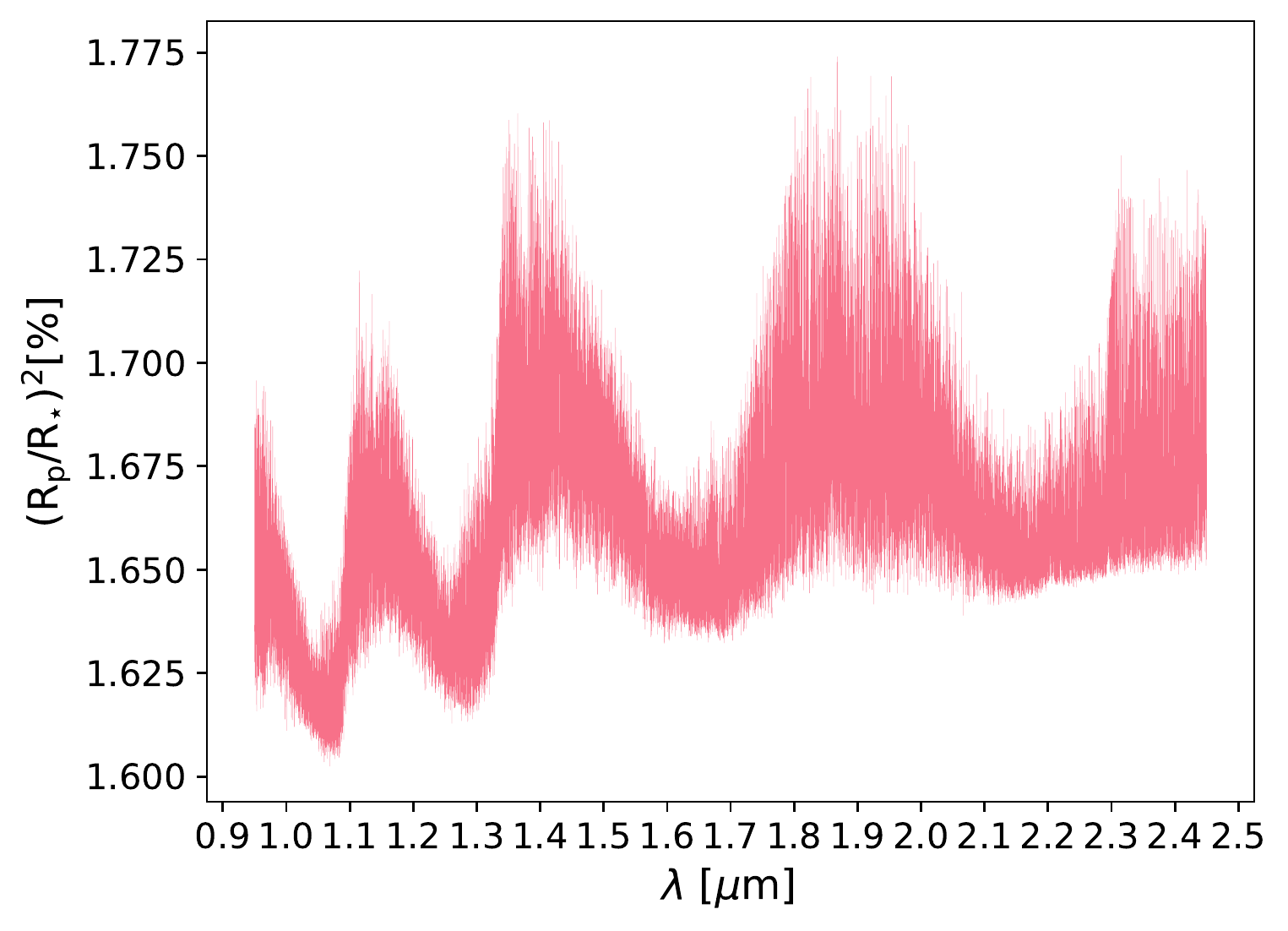}
   \includegraphics[width=0.49\textwidth]{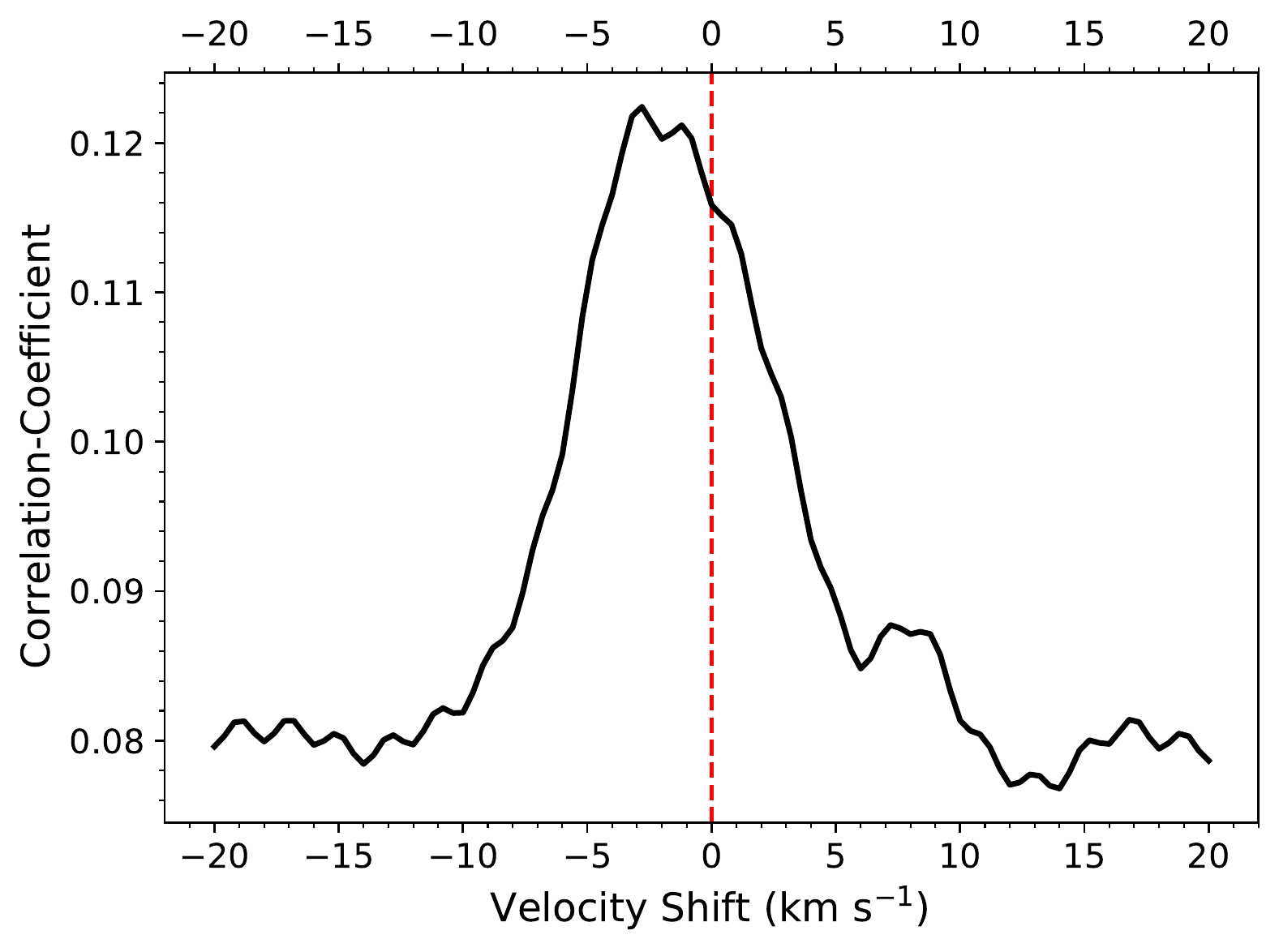}
   \caption{Left: high resolution transmission spectra of the Exo-FMS HD 209458b GCM results for the GIANO-B wavelength range (0.9-2.5$\mu$m)
   Right: cross-correlation result showing the net blueshift of the wind Doppler signature.}
   \label{fig:Exo-FMS}
\end{figure*}

For this demonstration we use output from the HD 209458b Exo-FMS GCM results \citep{Lee2021}.
We take the correlated-k model from \citet{Lee2021}.
We post-process this simulation to produce high-resolution transmission spectra across the GIANO-B wavelength range ($\approx$ 0.95-2.5 $\mu$m) \citep[e.g.][]{Giacobbe2021}.
We use a resolution of R = 150,000 (for a total of 142108 wavelength points) and produce spectra with and without Doppler and orbital shifting assuming the planet is at an orbital phase of zero.
The Doppler shifted spectrum across the wavelength range is shown on the LHS of Fig. \ref{fig:Exo-FMS}.

We then perform a cross-correlation analysis using the Doppler shifted spectrum as the mock observable and the non-Doppler shifted spectrum as the template.
The result of this is shown on the RHS of Fig. \ref{fig:Exo-FMS}.
A clear net blueshift of $\approx$3 km s$^{-1}$ can be seen in the peak of the CCF function, corresponding to the wind speeds found in the simulation \citep{Lee2021}.

This simple demonstration shows that gCMCRT is suitable for high-resolution transmission spectra modelling studies similar to \citet{Kempton2012, Flowers2019} and \citet{Savel2021}.
\citet{Wardenier2021} produced a detailed analysis of WASP-76b Fe depletion \citep{Ehrenreich2020, Kesseli2021} as a function of phase by post-processing a SPARC/MITgcm WASP-76b model using the CPU version of gCMCRT.

\section{Discussion}
\label{sec:disc}

With this new version of gCMCRT we have greatly extended the capabilities for RT modelling 3D planets at high spectral resolution.
Most 3D high resolution modelling studies to date \citep[e.g.][]{Kempton2012,Showman2013c,Flowers2019,Beltz2021,Harada2021,Wardenier2021} have focused on narrow wavelength bands, with the exception of \citet{Savel2021}.
We have shown gCMCRT can efficiently model across the IGRINS wavelength range ($\approx$ 1.45-2.45 $\mu$m) and GIANO-B ($\approx$ 0.95-2.5 $\mu$m) at moderately high resolutions of R = 135,000 and R = 150,000 respectively, with the number of discrete wavelength points numbering in the 100000s, suitable for cross-correlation of model results to elucidate physical processes.
This enables a much more detailed physical interpretation of the results of contemporary high-resolution studies, spreading across many more lines and probing the features of more species.

In Section \ref{sec:hiem} we performed mock cross-correlation detections of H2O and CO.
This would be relatively simple to extend to other species (such as HCN) by making template spectra and performing the same cross-correlation analysis.
In Section \ref{sec:hitrans} we only cross-correlated the gross model spectrum to find the net wind speed of the model.
By generating template spectra of each constituent species, it would be relatively simple to extend our approach here to model the detection of specific molecules and atoms directly from the GCM output, for example, producing predictions of the detected molecules in \citet{Giacobbe2021}.
Our high-resolution modelling capability provides a useful additional check to GCM models, where theorists and modellers can now directly compare their GCM output to the high-resolution observational data and see if their predictions about the 3D distribution of chemical species abundance, wind speeds, condensation and haze patterns are reasonable.

We specifically used outputs from many different modelling groups in this study.
This is to demonstrate the applicability of the model to many GCM output formats and to provide the community with baseline conversion scripts they can use to automate the processing of large grids of GCM data.

Our set-up and code of gCMCRT enables future implementations of other photon microphysics.
For example, heating and cooling rates of the atmosphere can be performed using the \citet{Lucy1999} method.
Well used CMCRT techniques from the astrophysical community such as photo-ionisation and dissociation \citep{Wood2013} can also be adapted for the planetary regime using this model as a baseline, enabling exploration of complex photon microphysics in 3D.
Expanding the model to performing 1D and using different geometries such as cartesian will further increase the flexibility of the model.

\section{Conclusions}
\label{sec:con}

gCMCRT is a GPU accelerated Monte Carlo Radiative Transfer code, suitable for global processing of 3D atmospheric data.
We develop standard albedo, emission and transmission spectra modes for typical use in producing observable properties from GCM model output.
We also develop a high-resolution spectral mode for emission and transmission spectra, able to capture the Doppler shifting of spectral lines due to rotation and atmospheric winds.

As a fully 3D model, gCMCRT avoids the biases and assumptions present when using 1D models to process 3D structures.
Atmospheric layers are weighted appropriately to their contribution to the end spectra

gCMCRT is well suited to performing the post-processing of large parameter GCM model grids starting to be performed by members of the community.
We developed simple pipelines that convert the 3D GCM structures from many well used GCMs in the community to the gCMCRT format, interpolate chemical abundances (if needed) and perform the required spectra calculation.
We include the ability to convert opacity tables from ExoMol to the gCMCRT format, enabling greater inter-comparison with well used 1D models.

The high-resolution spectra modes of gCMCRT provide an additional highly useful capability for 3D modellers to directly compare output to high-resolution spectral data.
Simulated wind speed, molecular detections and other variables can be calculated and compared to the observations through mock cross-correlation analysis of the models.

\subsection*{Support}

gCMCRT is provided as open-source software on the lead author's GitHub:
\url{https://github.com/ELeeAstro}

K-tables derived from HELIOS-k calculations for various gas species are also provided as download links on the lead author's GitHub.
These k-tables follow the CMCRT k-table format, and are at a resolution of R = 100 between 0.3 and 30 $\mu$m, suitable for modelling JWST data.
Additional species k-tables and cross-section tables can be created by the lead author upon request.
Python conversion scripts to the CMCRT table format are provided for the ARCiS, petitCODE and TauREx opacity tables from ExoMol.
These are required since gCMCRT does not use external packages by design.

Python preparation scripts with the correct data structures for gCMCRT are provided for 3D GCM input.
Custom python scripts are provided to aid conversion of the \citet{Rauscher2010,Rauscher2012,Rauscher2013} model, ExoRad, SPARC/MITgcm, UK Met-Office UM, THOR and Exo-FMS GCM output to the gCMCRT format.

Some CE abundance tables and python scripts are included to interpolate GGchem CE grid results to the GCM grid.

More detailed theory and documentation on the MCRT aspects are provided on the GitHub page.

The older CPU version, CMCRT, with similar capabilities as gCMCRT is available on a collaborative basis from the first author.

\begin{acknowledgments}
We thank M. Line for details and help with the IGRINS WASP-77Ab observations.
E.K.H. Lee is supported by the SNSF Ambizione Fellowship grant (\#193448).
J.P.W. sincerely acknowledges support from the Wolfson Harrison UK Research Council Physics Scholarship and the Science and Technology Facilities Council (STFC).
M.T.R. was supported by a European Research Council Consolidator Grant, under the European Union’s Horizons 2020 research and innovation program (\#723890).
Plots were produced using the community open-source Python packages Matplotlib \citep{Hunter2007}, SciPy \citep{Jones2001}, and AstroPy \citep{Astropy2018}.
The HPC support staff at AOPP, University of Oxford and University of Bern are highly acknowledged.
\end{acknowledgments}

\bibliography{bib2}{}
\bibliographystyle{aasjournal}



\end{document}